\documentclass[11pt]{article}

\usepackage[a4paper,margin=1in]{geometry}

\usepackage[utf8]{inputenc}
\usepackage[T1]{fontenc}
\usepackage{lmodern}

\usepackage{amsmath,amsfonts,amssymb,amsthm,amsopn,amstext}

\usepackage{booktabs}
\usepackage{rotating}
\usepackage{lscape}
\usepackage{longtable}
\usepackage{float}
\usepackage{array}
\usepackage{caption}
\usepackage{subcaption}
\usepackage{capt-of}
\usepackage{adjustbox}
\usepackage{arydshln}                 

\usepackage{xcolor}
\usepackage{graphicx}

\usepackage[round,authoryear]{natbib}
\usepackage{url}
\PassOptionsToPackage{hyphens}{url}
\usepackage[hidelinks,breaklinks=true,pdfencoding=auto,unicode=false]{hyperref}

\usepackage[toc,page]{appendix}
\usepackage{setspace}
\usepackage{titlesec}
\titleformat{\section}{\normalfont\large\bfseries}{\thesection.}{0.5em}{}
\titleformat{\subsection}{\normalfont\normalsize\bfseries}{\thesubsection.}{0.5em}{}

\makeatletter
\DeclareUrlCommand\harvard@@url{\itshape}
\renewcommand{\harvardurl}{\textbf{URL:} \harvard@@url}
\makeatother

\begin{document}

\newif\ifanon
\anonfalse

\title{\textbf{A World of Ginis}}
\ifanon
  \author{}
  \date{}
\else
  \author{Lidia Ceriani\thanks{Department of Economics, University of Verona, Verona, Italy. ORCID: 0000-0002-0372-6072.} \and Paolo Verme\thanks{Department of Statistics, Alma Mater University of Bologna, via Zamboni 33, 40126 Bologna, Italy. Email: \texttt{paolo.verme@unibo.it}. ORCID: 0000-0002-8754-1138.}}
  \date{July 24, 2026}
\fi
\maketitle

\begin{abstract}
\noindent The Gini index remains the most important measure of economic inequality worldwide, and accurate estimates of this index are essential for effective public policies. Yet, Gini estimates for the same country and year vary considerably across data sources, a problem that remains largely unresolved. The paper reviews the largest global and regional databases providing Gini estimates, surveys the related literature, and constructs a unified dataset of 122,351 Gini observations spanning 222 countries and territories and 158 years, from 1867 to 2024. The analysis of this new dataset shows that income-based Ginis exceed consumption-based ones by 4.7 points on average globally, and by as much as 10 points in some regions, with these gaps widening over time. The gross--net income distinction and the use of alternative equivalence scales together with several other measurement choices add further systematic differences. Based on these findings, the paper provides correction factors that can be used to harmonise Ginis built on different welfare concepts. We further show that overall divergence across databases has grown only modestly since 1960, and mainly through the proliferation of databases rather than through genuine divergence among long-standing sources. Thus, improving on the existing discrepancies across Ginis globally is possible, but ultimately depends on database administrators disclosing full details of Gini construction and on users selecting Ginis built on comparable measures.

\bigskip
\noindent \textbf{Keywords:} Gini coefficient; welfare measurement; cross-country comparability; inequality databases; regression; income--consumption gap.

\medskip
\noindent \textbf{JEL Classification:} D31, D63, I31, O15, C81.
\end{abstract}

\thispagestyle{empty}
\clearpage
\setcounter{page}{1}

\section{Introduction}\label{sec:intro}

Over the past three decades, evidence has accumulated from a wide range of countries and contexts that inequality affects economic growth, political stability, social mobility, health outcomes, and the efficacy of public policy \citep{Milanovic2016, Chancel2022, Piketty2014}. The policy implications are significant. If inequality is rising, the case for redistributive policy is strengthened. If it is falling, understanding whether this is associated with an economic downturn or redistributive policies would be essential. If apparent trends are artefacts of measurement choices, conclusions about the effectiveness of existing policies may need revision. Whatever the trends, changes in economic inequality signal important shocks to societies that require a thorough understanding.

Addressing these issues requires consistent and comparable data. Over the past thirty years, a proliferation of databases has made inequality data more accessible than ever before. The UNU-WIDER WIID \citep{UNUWIDER_2025}, the World Bank's Poverty and Inequality Platform \citep{PIP_2024}, the Luxembourg Income Study \citep{LIS_2024}, the World Inequality Database \citep{BlanchetEtAl2024}, the Standardized World Income Inequality Database \citep{Solt_2020}, All the Ginis \citep{Milanovic_2019}, together with a growing number of regional databases covering Latin America, Europe, Asia, and Africa, collectively provide tens of thousands of Gini coefficient estimates for over two hundred countries and multiple decades.

Yet, the proliferation of databases has not resolved the fundamental challenge of cross-country comparability. Decades ago, \citet{DeiningerSquire1996} assembled what was then the largest cross-country inequality dataset and confronted a challenge that remains to this day. Gini estimates drawn from different sources, even for the same country and year, diverge substantially. \citet{AtkinsonBrandolini2001} documented these problems systematically for OECD countries, showing that secondary dataset comparisons could mislead rather than inform analysis when the underlying concepts were not carefully harmonised. More recently, \citet{Jenkins2015} provided a comprehensive critical assessment of the two largest global secondary databases (WIID and SWIID), and \citet{FerreiraEtAl2015}, in their introduction to a special issue of the Journal of Economic Inequality devoted to cross-national inequality databases, called for more systematic assessment of the sources of discrepancy.

Building on \citet{Jenkins2015}'s assessment of the WIID and SWIID and on the empirical priorities articulated by \citet{FerreiraEtAl2015}, we extend the analysis to thirteen databases and provide econometric estimates of the magnitude and sources of discordance. We make four main contributions. First, we assemble a unified dataset of 122,351 Gini observations from 222 countries and territories covering the period 1867--2024 (to our knowledge, the largest consolidated collection of world Ginis to date) documenting coverage, genealogical relationships, and methodological characteristics across databases. Second, we conduct a systematic empirical analysis using four complementary strategies: within-country-year variability analysis, pairwise concordance matrices, OLS estimation of the income--consumption gap\footnote{In this paper, we use the terms consumption and expenditure interchangeably, the standard approach in the databases reviewed, none of which formally distinguishes between the two.} by region and income group, and a two-way fixed effects regression of Gini values on welfare concept and equivalence scale choices. Third, we provide what is, to our knowledge, the first systematic decomposition of the aggregate divergence trend, distinguishing database proliferation from genuine within-pair divergence. Fourth, we translate these findings into empirically grounded correction factors, with explicit documentation of the time periods and country groups for which they are most reliable.

Four main findings emerge beyond what the existing literature has already established. First, the welfare concept is the dominant source of cross-database discordance: income-based Ginis exceed consumption-based ones by an average of 4.7 Gini points globally, ranging from $+2.7$ points in Europe and Central Asia to $+10.2$ points in North America, and the gap widens monotonically as the level of development falls. The gross--net income distinction and the equivalence scale add further systematic differences, so that pooling Ginis built on different welfare concepts can materially bias cross-country comparisons. Second, these premia are not stable over time. The gross-income premium relative to consumption rose from 3.7 to 6.2 Gini points between the pre- and post-2000 periods, so correction factors estimated from historical data cannot be assumed to hold in more recent periods. Third, and as the most directly actionable contribution of the paper, we translate these regularities into a set of practical correction factors, disaggregated by World Bank region, income group, and welfare-concept dimension. These correction factors range from $+2.7$ Gini points (Europe and Central Asia) to $+10.2$ points (North America) for the consumption-to-income conversion and from $+0.3$ to $+3.7$ points for the net-to-gross conversion, giving empirical researchers a transparent first step toward combining Ginis drawn from incommensurable welfare concepts. Fourth, while cross-database disagreement for a given country-year is substantial, its average level has grown only modestly over time (about 0.03 Gini points per year since 1960), and a balanced-pair analysis and shift-share decomposition show that this modest growth is driven by the proliferation of databases rather than by long-standing databases diverging from each other.

The paper proceeds as follows. Section~\ref{sec:literature} provides background on the history of global inequality databases and the comparability challenge. Section~\ref{sec:databases} describes the thirteen databases and the construction and coverage of the unified dataset. Section~\ref{sec:discrepancy} analyses the sources of discrepancy. Section~\ref{sec:quantitative} presents the quantitative analysis. Section~\ref{sec:casestudies} provides country case studies. Section~\ref{sec:conclusion} concludes and draws implications for empirical research.

\section{Background}\label{sec:literature}

\subsection{An Overview of Global Inequality Databases}

The first systematic effort to compile a world database of Ginis emerged in the 1990s. \citet{DeiningerSquire1996} assembled a dataset of 693 high-quality Gini observations for 108 countries covering several decades, establishing quality criteria (nationally representative surveys, household income or expenditure as welfare concept, comprehensive coverage of the population) and making an early effort to document the welfare concepts underlying each observation. Their dataset was widely used in the growth-inequality literature of the late 1990s and also revealed how difficult it was to maintain conceptual consistency across countries and time periods.

To date, we can count at least five more global repositories of Gini indexes including the UNU-WIDER WIID, the Luxembourg Income Study, the World Inequality Database, the Standardized World Income Inequality Database (SWIID), and All the Ginis. The UNU-WIDER WIID, first released in 2000, expanded the Deininger--Squire approach, combining Ginis from dozens of sources into a single database with detailed documentation of welfare concepts \citep{UNUWIDER_2025}. The WIID now contains over 26,000 observations for 200 countries from 1867 to the present, making it the largest single repository of source-documented Gini data outside our newly assembled unified dataset (the SWIID contains marginally more observations, but its values are model-imputed rather than source-recorded; see Table~\ref{tab:databases}). The Luxembourg Income Study \citep{LIS_2024}, originally instituted to harmonise microdata on income and consumption across countries, allows researchers to estimate the Gini index for over 50 countries and over time with harmonised definitions. Although LIS covers far fewer countries and years than secondary databases, its estimates are widely regarded as the most internally consistent cross-national inequality data available. \citet{Solt_2020}'s Standardized World Income Inequality Database (SWIID) uses information from LIS to enhance the WIID and produce a new database. The author statistically estimates what each WIID country's Gini would look like under LIS-consistent concepts, maximising comparability. The World Inequality Database \citep{BlanchetEtAl2024} introduced the Distributional National Accounts (DINA) methodology \citep{PikettySaezZucman2018}, which anchors distributional estimates to national accounts totals. By combining household surveys with tax records and national accounts, the WID produces estimates consistent with macroeconomic aggregates but differing substantially from survey-based alternatives. Finally, \textit{All the Ginis} (ATG), created by Branko Milanovic, occupies a distinctive niche among global secondary databases. Originally assembled in 2004 and updated through 2019, ATG brings together 5,121 Gini observations for 175 countries covering the period 1948--2017 and drawing exclusively on Ginis calculated from actual household surveys with no estimates derived from regressions or interpolation \citep{Milanovic_2019}. A key feature of this database is its careful coding of methodological characteristics. Each observation records the income-recipient unit, the welfare concept (income or consumption), and whether income is gross or net of taxes. This metadata architecture prefigures the kind of standardisation this paper argues the field still needs.

The regional databases that complement these global sources have evolved along different trajectories. CEPAL/ECLAC, SEDLAC, and the IDB's Soci\'{o}metro all provide income-based Ginis for Latin America using national household surveys, but differ in their definitions of household income, treatment of imputed rent and pensions, and geographic coverage. The Eurostat EU-SILC represents perhaps the most successful example of methodological harmonisation, providing consistent estimates for 37 countries under a regulatory framework that mandates specific variable definitions. Yet, even in this case, some components of the welfare aggregate (such as imputed rents) are still not harmonised across countries. The ADB's Key Indicators Database and the OECD's Income Distribution Database impose some harmonisation while leaving enough flexibility to national agencies with the result that cross-country comparability remains imperfect. Afristat, compiling official statistics from 20 West and Central African countries, is the least harmonised of the databases included in this analysis, with limited information about underlying welfare concepts.

This diversity of architectures --- from the fully harmonised microdata approach of LIS, through WID's survey-plus-administrative-data method, to the secondary compilations of WIID and SWIID, to the minimally standardised regional databases --- reflects genuine trade-offs between coverage and comparability. Databases prioritising coverage must accept methodological heterogeneity while those prioritising comparability must accept limited reach. Our unified dataset assembles all these approaches for direct comparison.

\subsection{The Comparability Challenge}

The construction of global repositories of Gini indexes has allowed scholars to provide new insights into global inequality trends. \citet{BourguignonMorrisson2002}, for example, documented a long historical rise in global inequality from 1820 to 1992. \citet{SalaIMartin2006} argued that global income inequality fell sharply from 1980 to 2000, driven by growth in China and India. \citet{Milanovic2016} showed that global inequality has declined slightly in recent decades, driven by growth in emerging economies, even as within-country inequality has risen in many places. This new global perspective on inequality fuelled, in turn, a reassessment of globalisation, international trade and aid in promoting global welfare. It also contributed to placing inequality centre stage among the statistics considered by the UN global sustainable development goals (SDGs).

Yet, comparing the Gini figures provided by these public repositories remains a major challenge. \citet{AtkinsonMicklewright1992} showed for the formerly socialist economies that meaningful comparison is impossible without source-level documentation of definitions and survey instruments. \citet{GottschalkSmeeding1997}, surveying the early LIS evidence, demonstrated that harmonised microdata can overturn cross-country inequality rankings based on unadjusted national sources, a warning echoed for rich countries by \citet{BrandoliniSmeeding2009}. \citet{Deaton2005} documented the systematic and growing divergence between survey-based and national-accounts-based measures of average living standards, anticipating the debate that now surrounds the WID's Distributional National Accounts: \citet{PikettySaezZucman2018} and \citet{AutenSplinter2024} reach sharply different conclusions about the trend in US top income shares from the \emph{same} administrative tax data, illustrating that methodological choices can dominate measurement even when the underlying data are held fixed. The exchange between \citet{Jenkins2015} and \citet{Solt2015} crystallised the central trade-off among secondary databases, between the source-level documentation of the WIID and the imputation-based comparability of the SWIID. 

By carefully looking at Gini constructions, it is possible to pinpoint more precisely some major sources of discrepancies. A first source is the welfare concept. It is well known that income Ginis exceed consumption Ginis for the same country-year, and that switching between the two materially alters inferences about global distributional dynamics \citep{AnandSegal2008, Ravallion2014}, yet the magnitude and cross-country variation of this gap has not been systematically documented, and \citet{AtkinsonBrandolini2001} showed that simple controls for welfare concept are insufficient to equate even OECD estimates across databases. A second source is the under-measurement of top incomes. The World Inequality Report \citep{Chancel2022} and \citet{ChancelPiketty_2021} report that top-income shares are substantially higher than survey data suggest, and \citet{HlasnyVerme2018} quantified the upward bias in Gini estimates attributable to top-income non-response, a bias that varies significantly across countries, years, and types of data. \citet{Lustig2019} surveys the parallel literature on the ``missing rich'' in household surveys and shows that corrections for under-coverage at the top are themselves highly method-sensitive. A third source lies in the construction of the secondary databases themselves. \citet{Jenkins2015} provided a comprehensive critical assessment of the WIID and SWIID, identifying important differences in their construction, and \citet{FerreiraEtAl2015} noted that ``different databases often give very different pictures of inequality trends'' and called for more systematic assessment of discrepancies. 

Table~\ref{tab:priorstudies} summarises this literature and locates our contribution within it. Relative to these studies, we extend the assessment in scope (thirteen global and regional databases analysed jointly), in method (genealogy mapping, balanced-pair divergence analysis, and a fixed-effects regression), and in practical output (estimated correction factors by region and income group). The paper will also depart from previous contributions by describing, discussing, and quantifying a much larger set of potential measurement differences across Ginis. Moreover, it will show that comparability is a moving target. Databases are periodically revised, so the figures downloaded today may not match those used in earlier studies, an issue that hampers research reproducibility.

\begin{center}
[Table~\ref{tab:priorstudies} about here]
\end{center}

\section{A Unified Dataset}\label{sec:databases}\label{sec:unified}

For the purpose of clearly identifying the sources of discrepancies across Ginis published in cross-country databases, we pull together the data from thirteen databases, distinguishing between global and regional scope and between primary and secondary data sources. Primary sources compute inequality indicators directly from underlying microdata (household surveys or administrative records) while secondary sources collect indicators from other studies, reports, or databases.

Among global databases, the primary sources are LIS (Luxembourg Income Study), the World Bank Poverty and Inequality Platform (PIP), and the World Inequality Database (WID); the secondary sources are the UNU-WIDER World Income Inequality Database (WIID), \textit{All the Ginis} (ATG), and the Standardized World Income Inequality Database (SWIID). Among regional databases, the primary sources are CEPAL/ECLAC, SEDLAC, and IDB (all covering Latin America and the Caribbean) and Eurostat (Europe); the secondary sources are the OECD Income Distribution Database (high-income countries), the ADB Key Indicators Database (Asia-Pacific), and Afristat (West and Central Africa).

Table~\ref{tab:databases} summarises the key methodological characteristics of each database. The databases span very different observation sets: from the IDB's 45,656 observations (covering 26 Latin American and Caribbean countries over the period 1970--2023 with extensive sub-group disaggregations) to Afristat's 84 observations (covering 20 African countries since 1994). Global secondary databases (WIID, SWIID) each contain around 26,000 observations for roughly 200 countries. Primary global databases range from 930 observations (LIS) to 10,419 (WID).

\begin{center}
[Table~\ref{tab:databases} about here]
\end{center}

We append the thirteen databases by country (using ISO 3-letter country codes), year, and welfare concept. Where databases provide multiple Gini estimates for the same country-year (e.g., the WIID, which aggregates from multiple sources, or Eurostat, which provides three distinct income concepts), all observations are retained. The resulting unified dataset contains 122,351 observations spanning 222 countries and territories from 1867 to 2024.

A key task in the merging process is the construction of harmonised indicator variables for welfare concept, sub-metric type, and equivalence scale, which we derive from the metadata provided by each database and from direct inspection of the underlying survey documentation where available. These variables are necessarily imperfect, given the limited metadata provided by some databases, but they enable the quantitative analysis in Section~\ref{sec:quantitative}.

\begin{figure}[htbp]
{\centering \includegraphics[width=\linewidth,keepaspectratio]{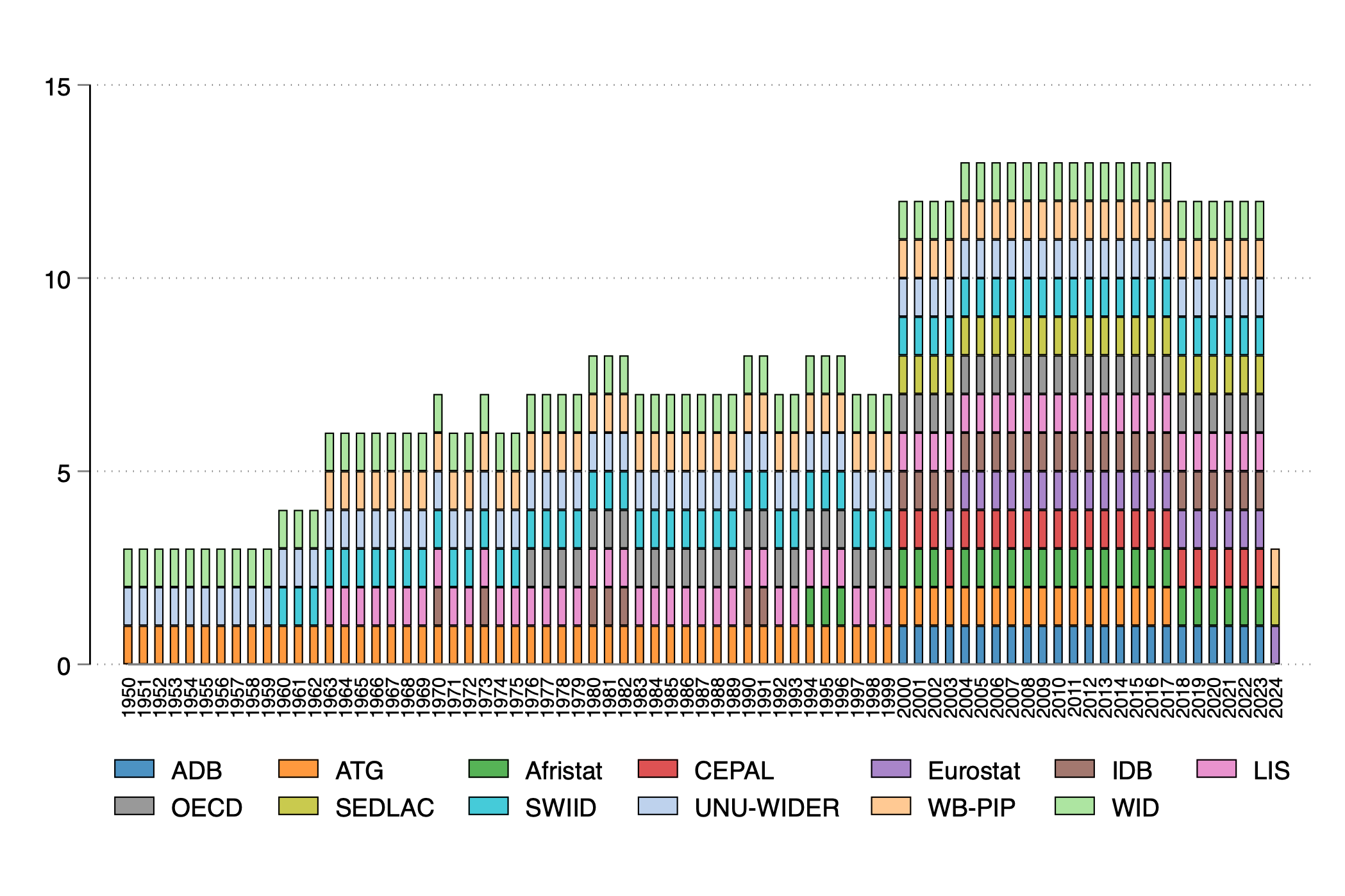}}
\caption{Number of Databases with Coverage per Year}
\label{fig:datasets}
\end{figure}

Figure~\ref{fig:datasets} shows the number of databases contributing coverage for each year since 1950 (the dataset extends back to 1867, but pre-1950 coverage is limited to one or two sources and is omitted from the figure for readability). Coverage is sparse before 1960, with the UNU-WIDER WIID providing the only observations before 1900 and only a handful of databases covering the 1900--1960 period. From 1960 onwards, the global secondary databases (WIID, SWIID) begin to provide substantial coverage, and primary databases expand rapidly after 1990. The densest coverage period is 2004--2017, which is covered by all thirteen databases. The number of distinct countries with Gini observations is below 40 before 1980; it exceeds 200 from 1980 onwards, when the WID's comprehensive country panel begins, while survey-based coverage excluding the WID rises above 100 countries only after 2000. The most recent years show some decline, reflecting publication lags in survey-based data.

\begin{center}
[Figure~\ref{fig:datasets} about here]
\end{center}

\begin{figure}[htbp]
{\centering \includegraphics[width=\linewidth,keepaspectratio]{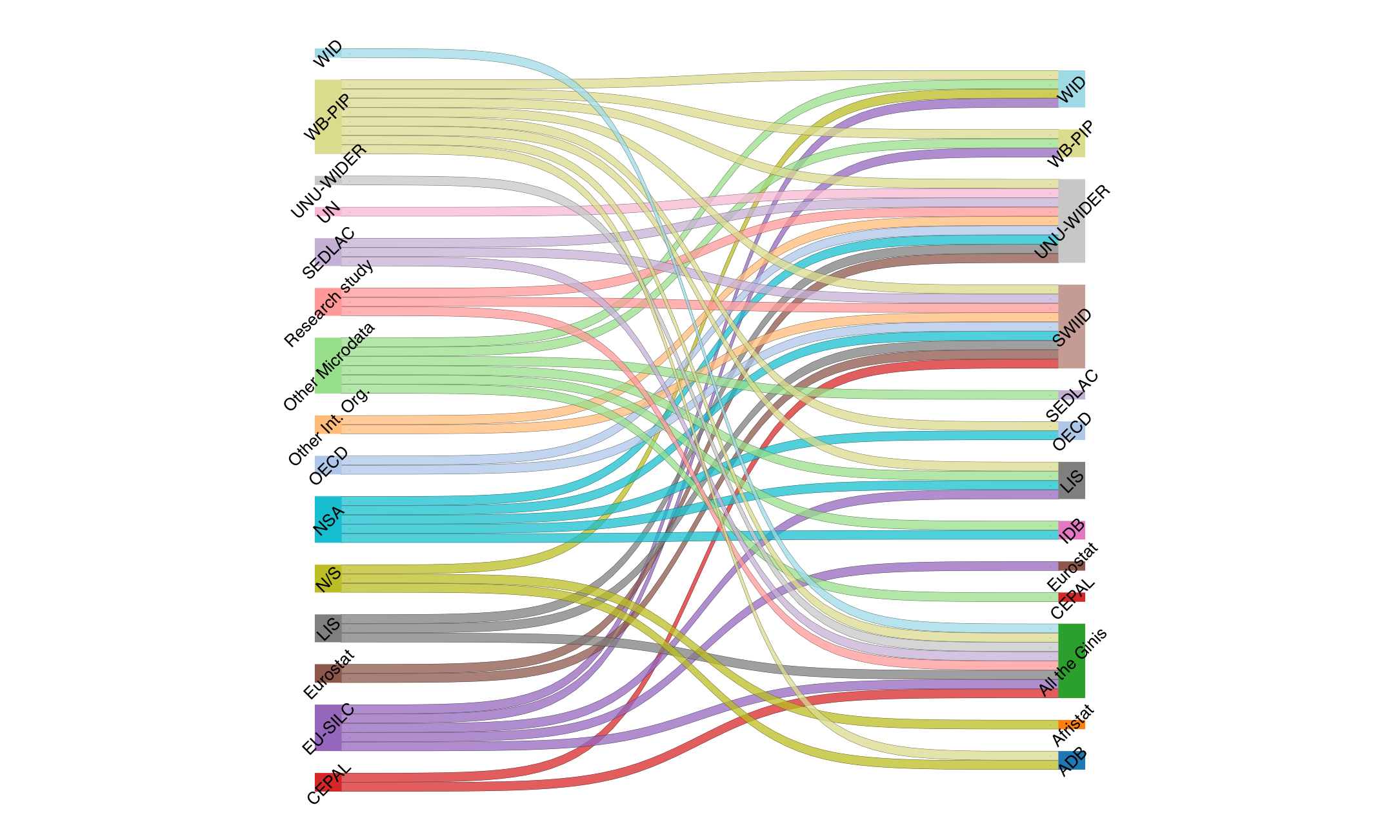}}
\caption{Genealogical Relationships Across Databases (Sankey Diagram). Left: data origins. Right: destination databases. All links are drawn at equal width: the diagram maps the existence of relationships between sources and databases, not their volume.}
\label{fig:sources}
\end{figure}

Observations from secondary databases are not statistically independent of observations from primary databases as they are often derived from them. Figure~\ref{fig:sources} maps the genealogical relationships between databases using a Sankey diagram. On the left are data origins (national statistical agencies, primary surveys, and existing databases) and on the right are destination databases. The global secondary databases (WIID, SWIID, and ATG) draw on the broadest range of sources, and their source sets overlap substantially. National statistical agencies are the ultimate origin of virtually all observations, feeding into primary databases (LIS, WB-PIP, WID, CEPAL, SEDLAC, IDB, Eurostat), which in turn feed into secondary databases. This genealogical structure has the important implication of non-independence. Any analysis that treats all 122,351 observations as independent is therefore flawed. Our analysis addresses this by focusing on pairwise and within-country-year comparisons rather than pooled regressions.

\begin{center}
[Figure~\ref{fig:sources} about here]
\end{center}

Tables~\ref{tab:databases} and \ref{tab:coverage_income} provide some important details. Table~\ref{tab:databases} reveals stark differences in mean Gini values across databases. The WID reports the highest mean (55.5), reflecting both its anchoring to national accounts (which estimates top incomes not captured in surveys) and the fact that it covers all countries, many of which are very unequal. The OECD IDD reports the lowest mean (31.9), reflecting its exclusive focus on high-income OECD countries. Among databases with broad global coverage, mean Ginis range from 37.1 (UNU-WIDER) to 39.5 (SWIID). The systematic differences in means across databases, which could reach 24 percentage points between WID and OECD, underscore the importance of understanding what each database measures. Table~\ref{tab:coverage_income} shows coverage by World Bank income group. The distribution is highly uneven. High-income countries are over-represented, with SWIID providing 1,763 country-year observations and UNU-WIDER providing 1,751, compared to 181 and 160 for low-income countries respectively. Regional databases specialise by design: Afristat covers almost exclusively low- and lower-middle-income countries; Eurostat covers almost exclusively high-income countries; the Latin American databases (CEPAL, SEDLAC, IDB) cover predominantly upper-middle-income countries. This uneven coverage has implications for analyses of the cross-country relationship between inequality and development. Studies that use databases with different income-group coverage will effectively be comparing different country samples, even if they claim to be studying the same phenomenon.


\begin{center}
[Table~\ref{tab:coverage_income} about here]
\end{center}

A broader point about the dataset design deserves explicit statement. The unified collection brings together survey-based Ginis, model-based estimates (SWIID), and national accounts-anchored measures (WID), objects that are not directly comparable. This is deliberate. The central claim of this paper is that these objects \emph{are} not comparable. In other words, researchers who draw on them interchangeably face a comparability problem of quantifiable magnitude. Demonstrating that claim requires assembling them in a single framework. We are therefore analysing non-comparable objects together precisely in order to measure the consequences of that non-comparability, not in spite of it.

\section{Sources of Discrepancy in Measured Inequality}\label{sec:discrepancy}

This section provides a systematic analysis of the main causes of variation in Gini values across databases, organised under four headings: survey design and implementation, welfare concept and sub-metric choices, reference unit and equivalence scale, and post-survey adjustments.

\subsection{Survey Design and Implementation}

All household surveys involve normative choices that affect distributional statistics. Among the most consequential for inequality measurement are the definition of the survey universe (which persons or households are included in the sampling frame), the reference period for income or consumption measurement, the treatment of irregular and seasonal incomes, and the treatment of non-response.

With respect to the survey universe, the choice of whether to include marginalised population groups, such as prisoners, institutional residents, homeless people, nomadic populations, refugees, and internally displaced persons, can affect measured inequality. In most countries, these populations are a small share of the total and their exclusion has little effect. But in countries with large displaced populations (Lebanon during and after the Syrian refugee crisis, Colombia during periods of internal displacement, or Iraq after 2003), or large populations of homeless people and prisoners (such as the US), exclusion of these groups can substantially understate inequality. Overall, these populations tend to be distributed towards the bottom of the income distribution and, as such, may weigh more on inequality measures than observations located around the mean or median values, although the Gini index may be more robust than other measures to extreme observations.

The reference period for income measurement is another consequential design choice. Annual income provides a comprehensive measure but requires recall of income received throughout the year, which is known to introduce systematic underreporting \citep{DeatonZaidi2002}. Monthly income is more reliably recalled but fails to capture income volatility and seasonal fluctuations. These differences in reference period produce structurally different income distributions and thus different Ginis \citep{BeegleEtAl2012}. Surveys that measure consumption over shorter reference periods (often two weeks for food, one month for non-food items) tend to yield more stable aggregate measures than those measuring annual income, which partly explains why consumption-based inequality measures are less volatile over time.

Several implementation factors affect the accuracy of measured distributions independently of design choices. Measurement error including systematic underreporting of income, particularly at the top of the distribution is one of the best-documented features of household income surveys \citep{HlasnyVerme2018}. Unit non-response (entire households refusing to participate) and item non-response (households participating but leaving particular income components blank) create further problems, and both types of non-response are concentrated at the top and bottom of the income distribution.

The Missing Not At Random (MNAR) character of top-income non-response is particularly problematic. If the high-income households systematically refuse to report their incomes, the resulting survey distribution will understate inequality. The magnitude of this bias depends on the fraction of top incomes that are missing and on how missing data are handled in post-survey processing. In some countries, top-income non-response rates can exceed 10 percent of high-income households, and the resulting bias in the Gini may be several percentage points \citep{HlasnyVerme2022}.

\subsection{Welfare Concept and Sub-metric Choices}

One important choice in the construction of the Gini is the choice of welfare metric, mostly the choice between income or consumption. This choice is not merely technical but reflects fundamentally different theories of welfare and fundamentally different possibilities for data collection in different country contexts.

From a welfare economics perspective, consumption has some theoretical advantages. It is a better proxy for permanent income, less affected by transitory income shocks, and more directly linked to living standards. From a practical perspective, consumption is more reliably measured in surveys in low-income countries where income is largely informal and irregular whereas income is more reliable in high-income countries with formal labour markets and standardised income flows \citep{DeatonZaidi2002, Ravallion2014}.

\begin{figure}[htbp]
  \begin{subfigure}[b]{0.49\linewidth}
    \centering
    \includegraphics[width=\linewidth,keepaspectratio]{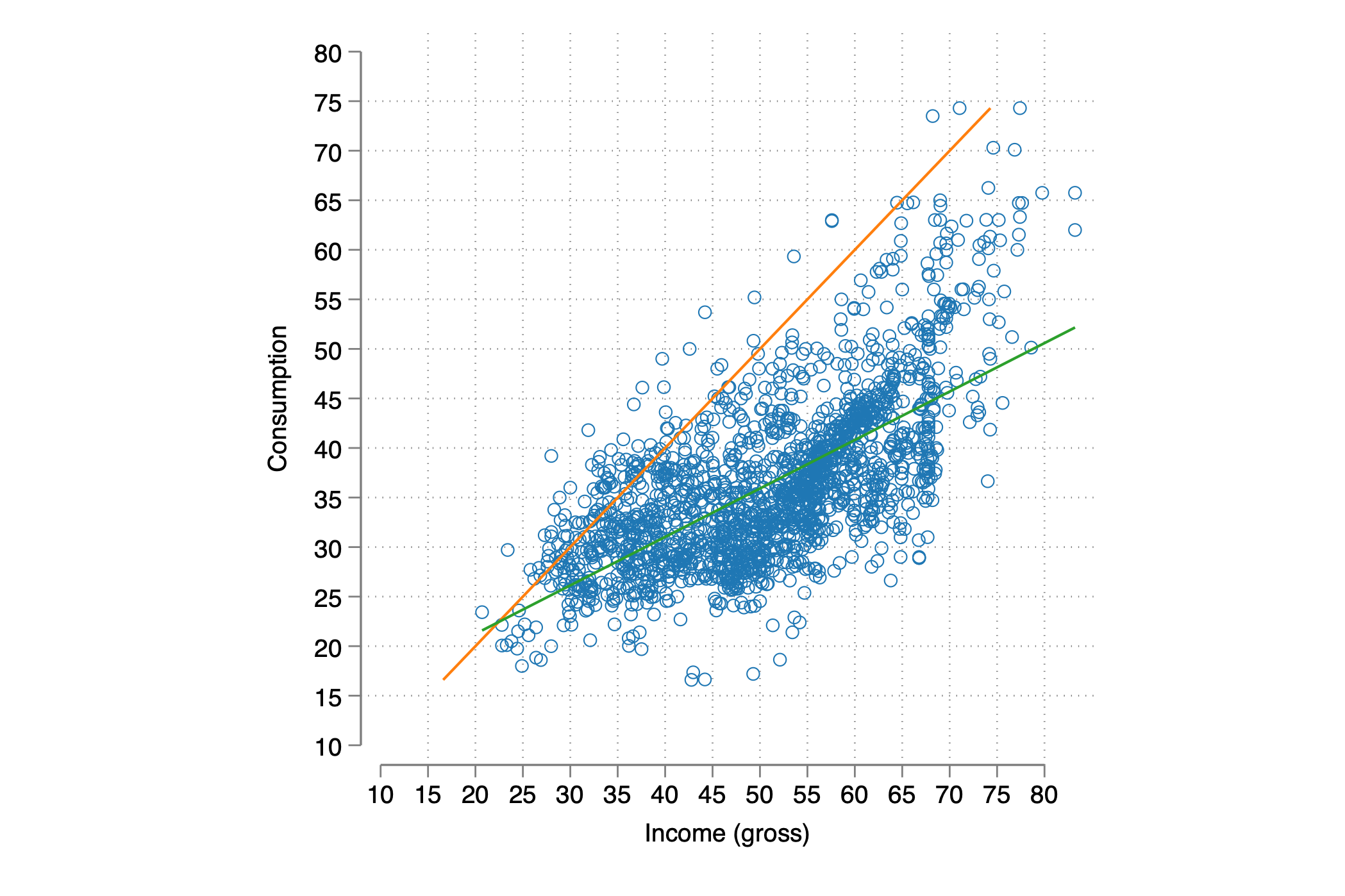}
    \caption{Gross income vs.\ consumption}
    \label{fig:incgross}
  \end{subfigure}
  \hfill
  \begin{subfigure}[b]{0.49\linewidth}
    \centering
    \includegraphics[width=\linewidth,keepaspectratio]{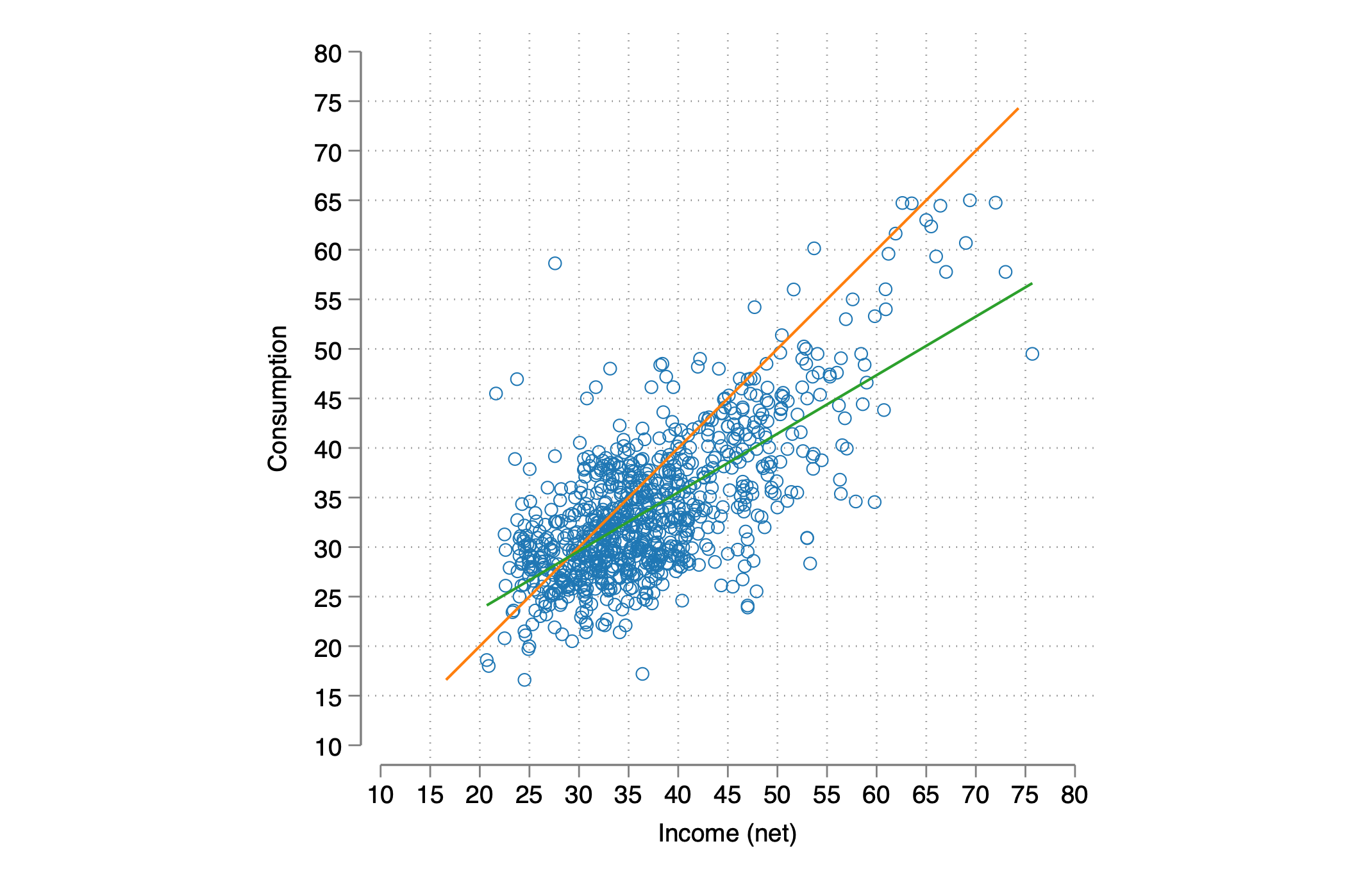}
    \caption{Net income vs.\ consumption}
    \label{fig:incnet}
  \end{subfigure}
  \caption{Income Ginis vs.\ Consumption Ginis for matched country-year observations. Each point is one country-year (consumption on the vertical axis, income on the horizontal axis); the diagonal represents perfect concordance. Panel~(a) shows gross income; Panel~(b) shows net disposable income. In both cases the points lie systematically below the 45-degree line --- income Ginis exceed consumption Ginis --- with the gap larger and more dispersed for gross income.}
  \label{fig:inc_cons_scatter}
\end{figure}

Income Ginis are consistently higher than consumption Ginis. Figure~\ref{fig:inc_cons_scatter} compares, for all country-year pairs where both an income Gini and a consumption Gini are available, the values of the two measures. Both panels show a clear, systematic departure from the 45-degree line, with points mostly below the line (consumption on the vertical axis), indicating that income Ginis tend to be larger than consumption Ginis. For gross income (Panel~a), the deviation is especially pronounced, with the scatter cloud lying well below the diagonal. For net income (Panel~b), the gap is smaller but still considerable, with individual discrepancies approaching 30 Gini points.

\begin{center}
[Figure~\ref{fig:inc_cons_scatter} about here]
\end{center}

\begin{figure}[htbp]
{\centering \includegraphics[width=\linewidth,keepaspectratio]{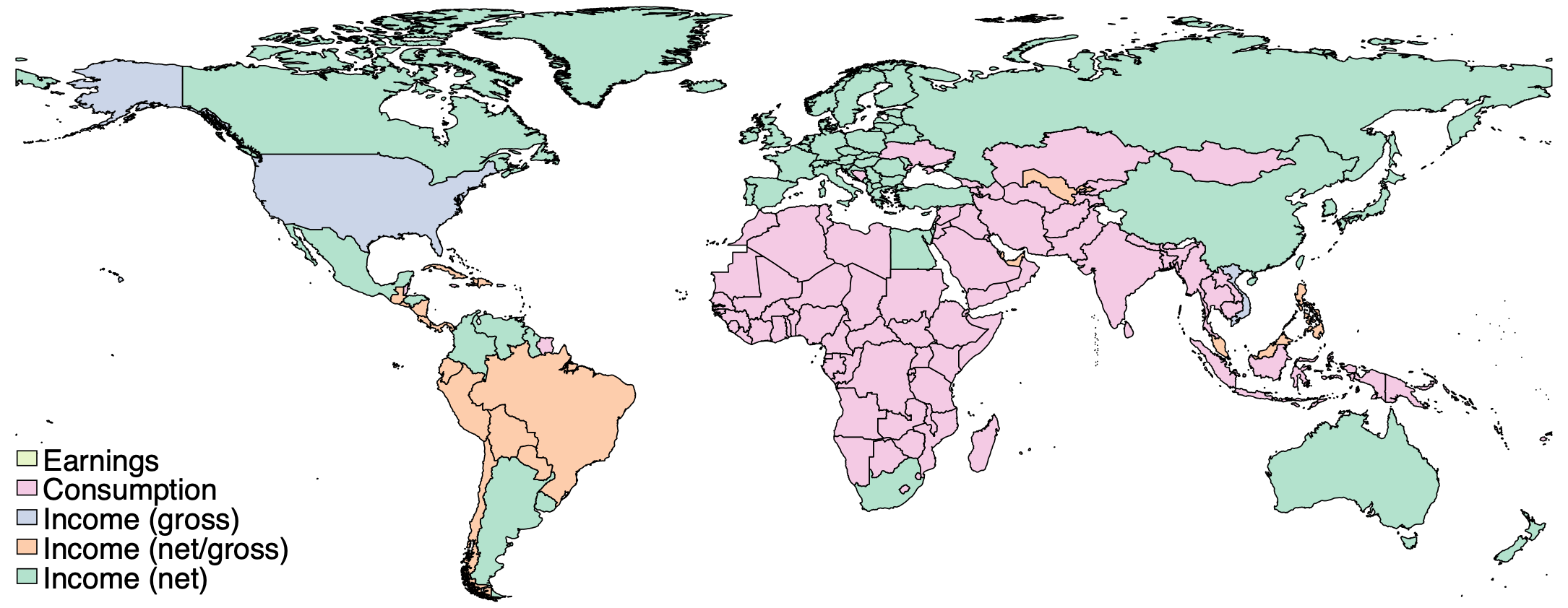}}
\caption{Countries' Prevalent Welfare Metric (most recent observation, most frequent definition). For each country, the map retains the most recent year with available data; where several databases, or several definitions within a database, report for that year, the modal (most frequent) welfare concept is shown.}
\label{fig:maptype}
\end{figure}

Figure~\ref{fig:maptype} shows the world map coloured by the prevalent welfare metric in each country's most recent database observation. The geographical divide is stark. Sub-Saharan Africa and South Asia are almost uniformly consumption-based, the Americas, Europe, and much of East Asia are predominantly income-based, and the Middle East and North Africa are largely consumption-based, with only a few exceptions. This structural divide means that any cross-regional comparison that mixes income and consumption Ginis without correction will systematically bias conclusions about the level and gradient of inequality across the development spectrum.

\begin{center}
[Figure~\ref{fig:maptype} about here]
\end{center}

Within any welfare metric, there is further heterogeneity arising from how the metric is constructed. For income, the most important distinctions concern whether the measure is gross or net of taxes and social contributions, whether it includes cash social transfers (unemployment benefits, family allowances, social assistance) and pensions (sometimes classified as deferred wages rather than transfers), and whether it captures imputed rent from owner-occupied housing or non-monetary income such as home production and in-kind public services.

A particularly instructive case is the treatment of imputed rents, the in-kind flow of housing services that owner-occupiers derive from their dwellings. Because it accrues in kind rather than in cash, it sits awkwardly within surveys designed to capture monetary receipts, and the decision to include or omit it can materially alter the measured distribution. Imputed rent is at once a component of income (an implicit return on housing wealth) and of consumption (the value of housing services consumed), so its treatment bears directly on the comparability of income- and consumption-based estimates. The adjustment is rarely trivial: housing is most households' principal asset, and its imputed value is lumpy and unevenly distributed across owners, renters, and those in rent-free accommodation. The methodological choices involved---self-assessment, hedonic regression, capital-market returns, or user-cost approaches---can yield appreciably different welfare rankings and inequality levels \citep{BalcazarEtAl2017}, with effects large enough to alter the international ranking of countries \citep{CerianiOlivieriRanzani2023}.

\begin{figure}[htbp]
{\centering \includegraphics[width=\linewidth,keepaspectratio]{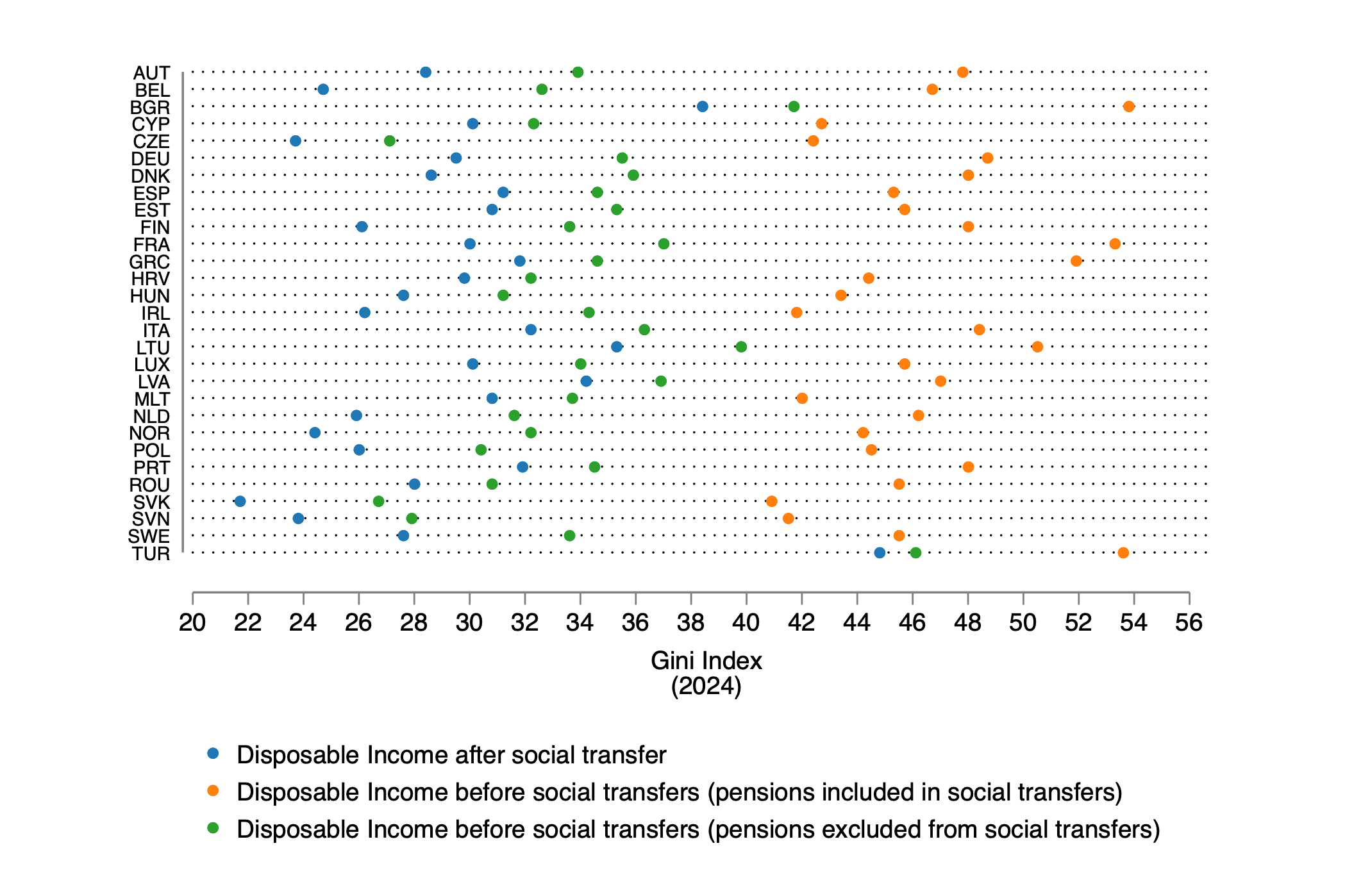}}
\caption{Ginis from Different Income Sub-Metric Definitions (Eurostat, 2024). The three Eurostat series correspond to: disposable equivalised income; disposable income before social transfers (pensions included); and disposable income before all social transfers.}
\label{fig:submetrics}
\end{figure}

Figure~\ref{fig:submetrics} illustrates the magnitude of these differences using Eurostat data, which is unique in reporting multiple income concept variants simultaneously for the same country and year. The three Gini variants provided by Eurostat (disposable equivalised income, disposable income before social transfers with pensions included, and disposable income before all transfers) differ by large amounts. When pensions are counted as transfers (and thus excluded), the pre-transfer Ginis are typically 10--20 Gini points higher than the post-transfer Ginis, with gaps exceeding 20 points in some countries; when pensions are retained as income, the gap narrows to roughly 3--8 Gini points. This reflects the substantial role of the welfare state in European countries in reducing inequality through taxes and transfers. A researcher who uses a pre-transfer Gini from one database and a post-transfer Gini from another will dramatically overstate the inequality differential between countries.

\begin{center}
[Figure~\ref{fig:submetrics} about here]
\end{center}

For consumption, the main sub-metric distinction is between a narrow ``food and non-food expenditure'' measure and a broader measure including imputed rent, home production, and durables. The broader measure tends to be more equal (lower Gini) because imputed rent and home production are relatively evenly distributed \citep{BalcazarEtAl2017}. While the importance of these differences may seem obvious to scholars working on these measures, the documentation that accompanies the databases reviewed in this paper rarely provides adequate information about which consumption items are included. A problem that equally applies to income.

\subsection{Reference Unit and Equivalence Scale}

Two further choices affect the shape of the distribution, the unit of observation (individual or household) and the equivalence scale (per capita, adult equivalent with various parameters, or unequalised household).

The choice between individual and household as the unit of observation is relatively straightforward. Individual-level measures count each person once and household-level measures count each household once, regardless of size. Since large households tend to have lower per-capita income, the household-based distribution is more compressed than the individual-based one, and the resulting Gini lower. However, not all databases report whether the Gini has been measured using households or individuals, which can result in very substantial differences across Ginis.

More complex is the choice of equivalence scale within individual-level measures. Three conventions are common: per capita (dividing total household income or consumption by the number of household members), the OECD-modified equivalence scale (assigning 1.0 to the first adult, 0.5 to each additional adult, and 0.3 to each child), and the square-root scale used by LIS and the OECD. Per-capita measures tend to produce higher Ginis because they do not account for economies of scale in household consumption whereas adult-equivalent measures partially adjust for scale economies.

\begin{figure}[htbp]
{\centering \includegraphics[width=\linewidth,keepaspectratio]{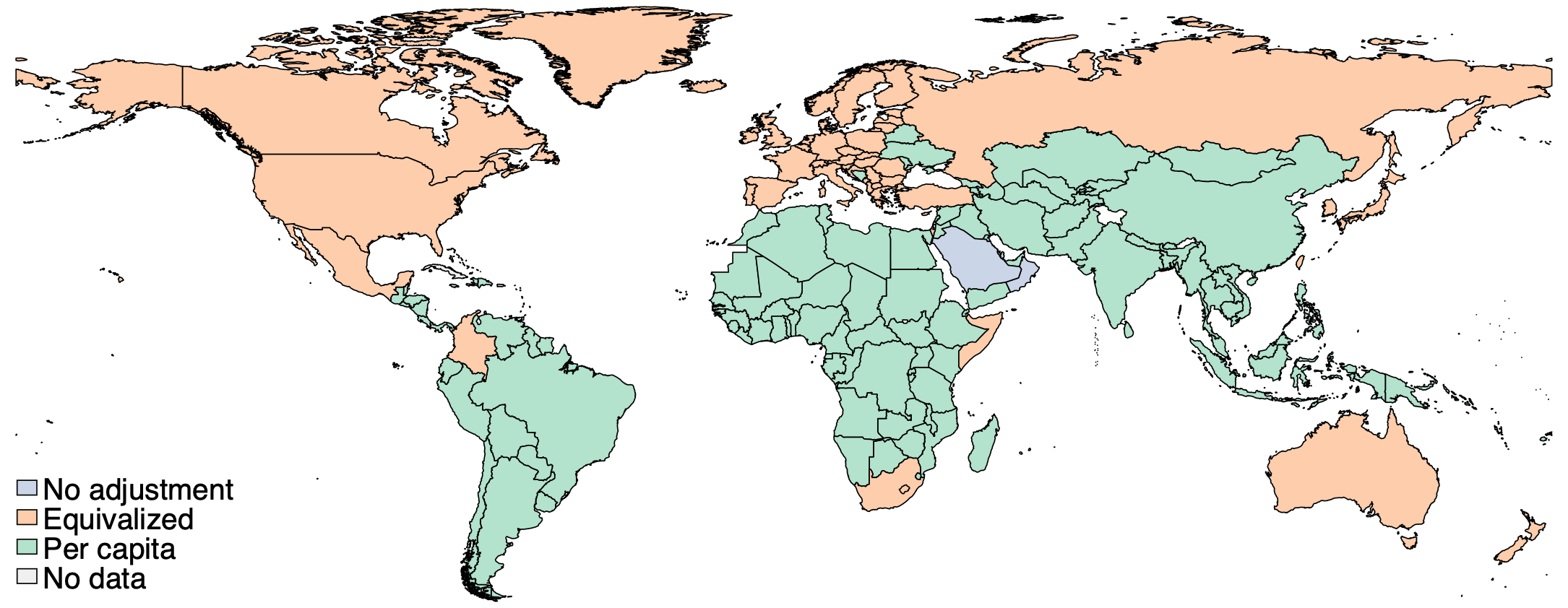}}
\caption{Countries' Prevalent Equivalence Scale (most recent observation, most frequent definition). For each country, the map retains the most recent year with available data; where several databases, or several definitions within a database, report for that year, the modal (most frequent) equivalence scale is shown.}
\label{fig:mapscale}
\end{figure}

\begin{figure}[htbp]
{\centering \includegraphics[width=\linewidth,keepaspectratio]{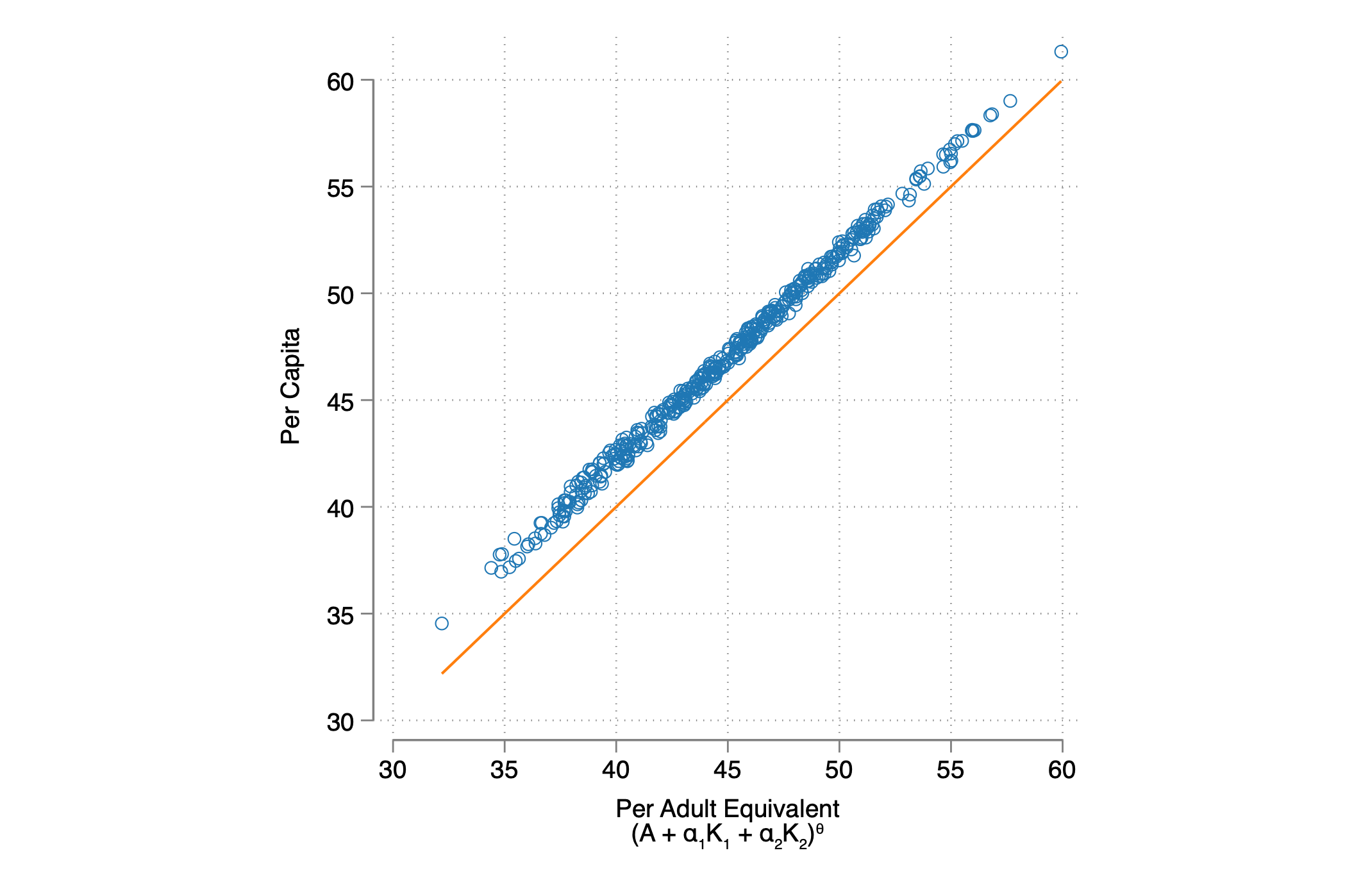}}
\caption{Per Capita vs.\ Adult Equivalent: Effect on the Gini (SEDLAC data). Each point is a country-year. The horizontal-axis expression is the SEDLAC adult-equivalence scale, $(A + \alpha_1 K_1 + \alpha_2 K_2)^{\theta}$, where $A$ is the number of adults, $K_1$ and $K_2$ are the numbers of younger and older children, and $\theta$ captures household economies of scale.}
\label{fig:scale}
\end{figure}

Figure~\ref{fig:mapscale} maps the world by prevalent equivalence scale in each country's most recent observation. Per-capita measures predominate in the Global South while adult-equivalent measures predominate in Europe and high-income countries generally. This geographical divide parallels the income--consumption divide and compounds comparability problems. Countries in the Global South are typically compared using consumption per capita, while countries in the Global North are compared using income per adult equivalent. Both differences push the Global South Ginis toward lower values relative to the Global North. Figure~\ref{fig:scale} illustrates the same parameters for SEDLAC data and shows that moving from adult-equivalent to per-capita income raises the Gini by approximately 2--4 percentage points for Latin American countries. This is modest compared to the income--consumption gap but not negligible.

\begin{center}
[Figure~\ref{fig:mapscale} about here]
\end{center}

\begin{center}
[Figure~\ref{fig:scale} about here]
\end{center}

\subsection{Post-Survey Adjustments and Database Revisions}

Several post-survey adjustments such as anchoring data to national accounts, top-coding, and price adjustments can also affect the Gini.

The most far-reaching post-survey adjustment in the current global inequality literature is the anchoring of survey distributions to national accounts aggregates, as implemented in the WID's DINA methodology. The DINA framework argues that household surveys markedly undercount top incomes, because the very rich do not participate in surveys or underreport their incomes. It corrects for this by using a combination of tax records, national accounts, and survey data to produce a distribution consistent with the national accounts aggregate. The resulting distributions show considerably higher top income shares and Gini coefficients than unadjusted surveys. The validity of this correction is debated --- most visibly in the exchange between \citet{PikettySaezZucman2018} and \citet{AutenSplinter2024}, who derive sharply different US inequality trends from the same tax data --- but its effect on measured inequality is clear. The mean WID Gini (55.5) is higher than the mean for survey-based databases with comparable country coverage (37--39 for WIID, SWIID, and ATG). The difference is not merely a level difference but also changes trends. Some analyses using WID data show rising within-country inequality where survey-based analyses show stable or declining inequality.

A different but still important factor for the measurement of inequality is top-coding which is applied by many national statistical agencies to microdata before public release. This is the practice of replacing the highest income values with a common cap or with synthetic values, or swapping records across observations with the intent of protecting the identity of high-income respondents. Top-coding is therefore an anonymisation device, not a correction for non-response bias. However, it truncates or compresses the upper tail of the distribution introducing a downward bias in measured inequality that compounds the bias already induced by top-income non-response. \citet{HlasnyVerme2022} show that for the United States, the Gini computed from public-access top-coded data and the Gini estimated after correcting for top-coding diverge by as much as 10 percentage points.

Price adjustments, while rarely discussed by the documentation that accompanies the global databases, can significantly change the estimation of the Gini. When data are pooled across countries (as in the World Bank PIP) or across time and space (in most databases), price adjustments using PPP conversion factors or consumer price indices that adjust prices spatially or longitudinally are necessary. The choice of base year, price index, and geographic price deflators can affect measured inequality levels. This issue is especially pressing for the World Bank's global poverty and inequality estimates, which rely on PPP conversion factors that are periodically revised making prior estimates of the Ginis obsolete.

In fact, the Gini values in most databases are not fixed once published but are subject to periodic revision. The introduction of the World Bank's Poverty and Inequality Platform in 2022, for example, revised historical inequality and poverty estimates for many countries, in some cases altering trend estimates spanning decades. Each new round of the International Comparison Programme (ICP)\footnote{See https://www.worldbank.org/en/programs/icp for more details.} changes the PPP conversion factors used for global comparisons and the 2011 and 2017 ICP rounds each produced non-trivial revisions to cross-country poverty and inequality figures. The WID periodically revises its DINA methodology and incorporates new tax record data, altering historical Gini series even for countries with no change in underlying survey data. The WIID, SWIID, and ATG are updated at irregular intervals, and updates sometimes correct errors, change the weighting of underlying sources, or revise the metadata for individual observations. Again, much of this work is not described in detail in the respective database documentation, leaving users to wonder how comparable the published Ginis are.

These revisions evidently create a replication problem. A study published in, say, 2018 using one vintage of a database may not be replicable from the current version of that database, even if the original underlying surveys are unchanged. More fundamentally, two studies that use the same database but different vintage years (not an uncommon situation given the existing publication lags) may reach different empirical conclusions for reasons that have nothing to do with economics and everything to do with the revision history of the data.

We flag this as a distinct dimension of non-comparability because it operates at a different level from the within-database methodological issues discussed above. Even a researcher who carefully controls for welfare concept, equivalence scale, and survey design faces the additional challenge that the specific numbers in the database may change between the time of data download and the time of publication, or between their study and a study attempting replication. Advocating for versioned, citable database releases with stable digital object identifiers for each vintage is therefore an important complement to the methodological harmonisation agenda.

\section{A Quantitative Analysis of Cross-Database Divergence}\label{sec:quantitative}

We now quantify cross-database divergence using four complementary approaches: a simple analysis of the cross-database variability of the Gini over time, a Gini pairwise cross-database concordance analysis, an econometric analysis of the Gini looking at the income--consumption gap, and a regression of the Gini across welfare concepts, equivalence scales, and time. The analyses (Sections~\ref{subsec:variability}--\ref{subsec:metareg}) are based on a ``core'' sample restricted to nationally representative, total-population observations. This restriction excludes the World Inequality Database because WID observations carry a sub-national population-coverage code and do not record welfare-concept or equivalence-scale metadata in machine-readable form. WID is retained in the database overview of Section~\ref{sec:unified} (Table~\ref{tab:databases}) but excluded from the income-group coverage tabulation (Table~\ref{tab:coverage_income}).

\subsection{Cross-Database Variability and Its Temporal Trend}\label{subsec:variability}

Our first approach calculates the absolute range (maximum minus minimum Gini) and standard deviation of Gini values across all databases, for each country-year with at least two database observations. This measures the extent to which databases disagree about the level of inequality in a given country and year. Table~\ref{tab:variability_region} presents results by World Bank region (Panel A) and income group (Panel B). Across 3,419 country-year observations, the mean within-country-year range is 4.25 Gini points and the median is 3.27 points. The distribution of ranges is heavily right-skewed, with a maximum that reaches 43 points.

The heterogeneity across regions and income groups is striking. High-income countries, which are most extensively covered by multiple databases, show the highest variability (mean range 5.5 pp). This reflects the diversity of welfare concepts used across databases for these countries: income-based (LIS, OECD), consumption-based (WB-PIP), and mixed (WIID, ATG). Low-income countries, by contrast, have low variability (mean range 2.4 pp), reflecting their reliance on a smaller number of databases using similar consumption-based approaches. North America (mean range 6.4 pp) and Europe and Central Asia (5.3 pp) have the highest regional variability. Both are predominantly high-income regions with dense database coverage and wide diversity in income concepts. Latin America and the Caribbean has a maximum range of 43 points, the highest of any region, reflecting the presence of many databases with very different welfare concepts (with the income-based Latin American regional databases sitting alongside consumption-based and mixed-concept global sources).

\begin{center}
[Table~\ref{tab:variability_region} about here]
\end{center}

\begin{figure}[htbp]
{\centering \includegraphics[width=\linewidth,keepaspectratio]{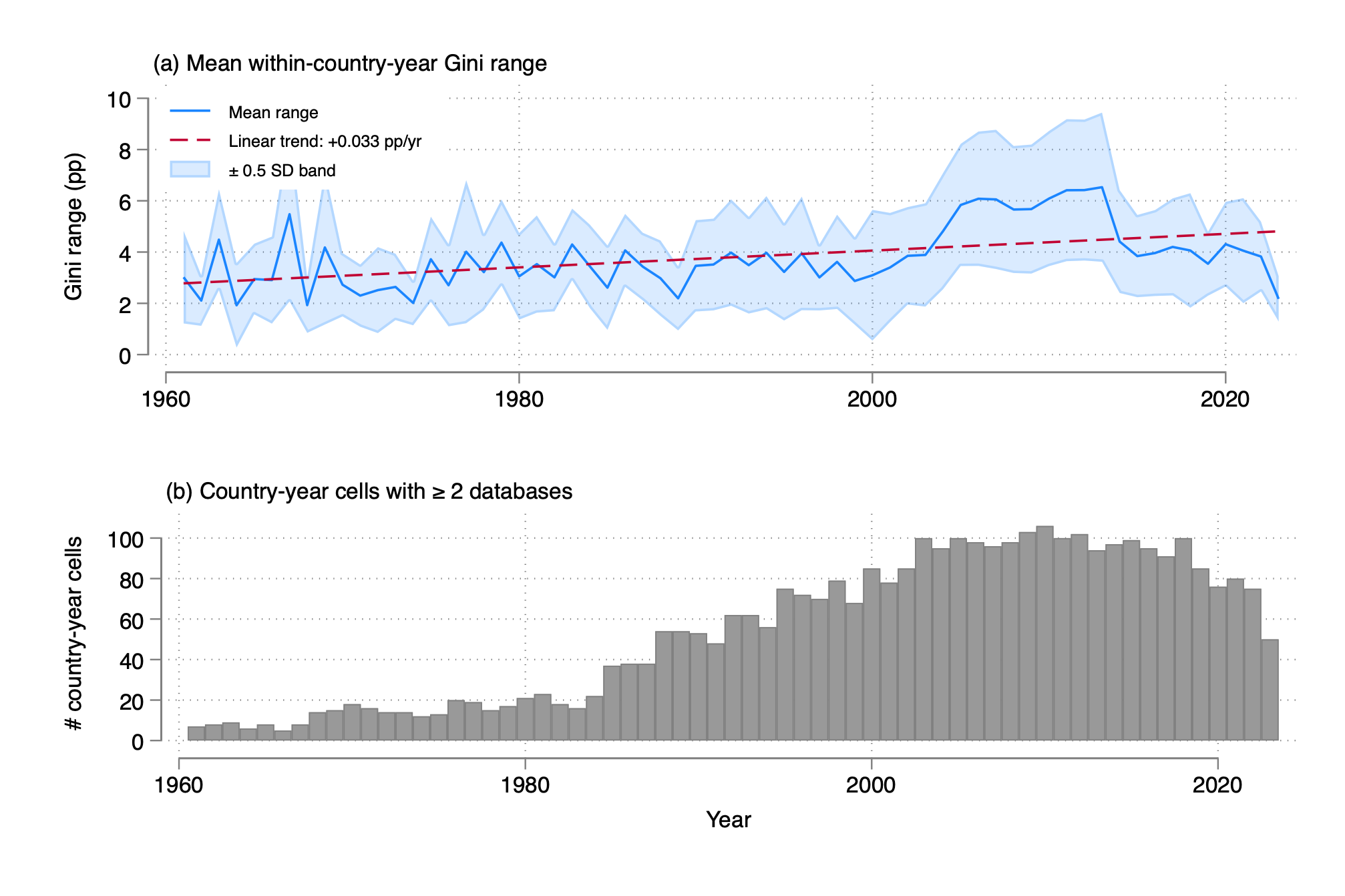}}
\caption{Cross-Database Gini Divergence Over Time. Panel (a): Mean within-country-year Gini range (in Gini points) for country-year cells covered by at least two databases, 1960--2023, with $\pm 0.5$ SD band and linear trend line. The trend coefficient is $+$0.033 Gini points per year. Panel (b): Number of country-year cells with $\geq 2$ database observations. Core sample (nationally representative, total-population observations); each database enters at its median Gini per country-year.}
\label{fig:divergence_trend}
\end{figure}

\label{subsec:timetrend}
A natural question is whether the degree of cross-database disagreement has increased over time or decreased as more databases have been created and updated. To investigate this question, we compute the mean within-country-year range for each year from 1960 to 2023 (restricting to years with at least five multi-database country-year observations) and plot the series alongside a simple linear trend. Figure~\ref{fig:divergence_trend} (Panel a) presents the aggregate results. The trend is positive but modest, with the mean cross-database range growing at $+0.033$ Gini points per year (Newey--West standard error $0.011$; 95\% confidence interval $[0.010, 0.055]$), so that the fitted trend rises by roughly two Gini points over the six decades of the sample. Panel~(b) reports the number of country-year cells with at least two database observations underlying each annual mean, a count that rises steadily over the period and confirms that the upward trend in Panel~(a) reflects a growing rather than a thinning sample of multi-database country-years.

\begin{center}
[Figure~\ref{fig:divergence_trend} about here]
\end{center}

\begin{figure}[htbp]
{\centering \includegraphics[width=\linewidth,keepaspectratio]{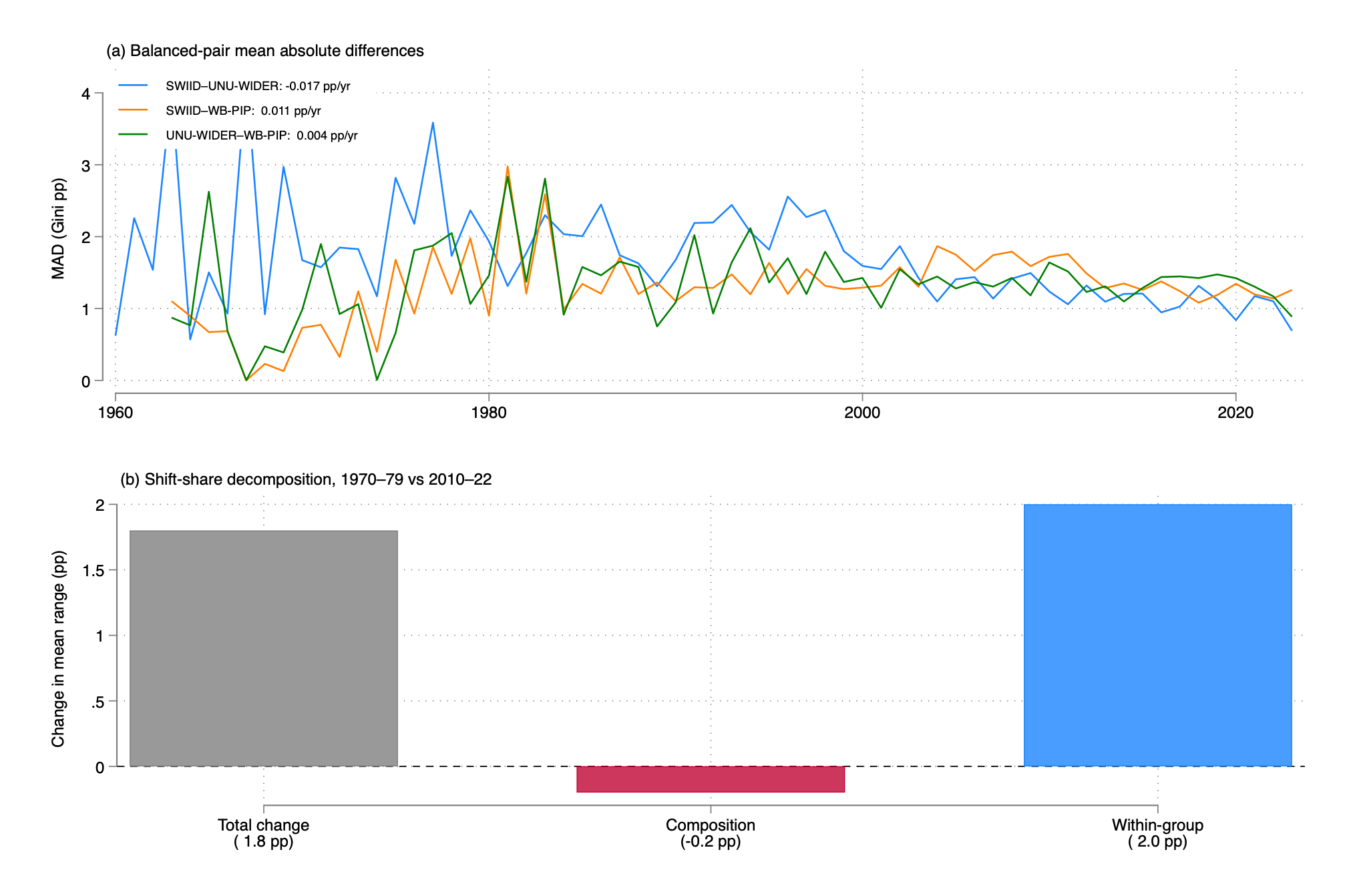}}
\caption{Decomposing the Aggregate Divergence Trend. Panel (a): Balanced-pair mean absolute differences (MAD) over time for three long-running database pairs; OLS trend slopes shown in the legend---all small in absolute value. Panel (b): Shift-share decomposition of the total change in mean within-country-year range between 1970--79 and 2010--22 ($+$1.8 pp) into composition (changing income-group weights; $-$0.2 pp) and within-group (changing ranges within income groups; $+$2.0 pp) components. The negative composition effect and negligible balanced-pair MAD trends jointly indicate that the aggregate trend is driven by more databases per country-year rather than by long-standing databases diverging from each other. Core sample (nationally representative, total-population observations).}
\label{fig:div_decomp}
\end{figure}

We then ask whether this aggregate trend reflects \emph{genuine within-pair divergence}, meaning existing databases disagreeing more with each other over time, or merely the \emph{intensive margin of database proliferation}, meaning more databases covering the same country-years mechanically producing larger observed ranges even if each pairwise relationship is unchanged. Answering this question requires moving from the aggregate range to two complementary analyses, reported in Figure~\ref{fig:div_decomp}.

The first temporal analysis is a balanced-pair analysis. For each of three long-running database pairs (SWIID--UNU-WIDER, SWIID--WB-PIP, and UNU-WIDER--WB-PIP) we compute the mean absolute difference (MAD) in matched country-year Gini values by year (Panel a). All three pairwise trends are small in absolute value. The SWIID--UNU-WIDER MAD declines at $-$0.017 pp/yr, the UNU-WIDER--WB-PIP MAD shows a negligible increase of $+$0.004 pp/yr, and the SWIID--WB-PIP MAD rises at $+$0.011 pp/yr. In no case does the time-averaged MAD exceed about 2 Gini points, and in no case is there evidence that these long-standing databases are becoming substantially less consistent with each other over time.

The second temporal analysis is a shift-share decomposition of the aggregate range trend (Panel b). We decompose the total change in mean within-country-year range between the 1970--79 base period and the 2010--22 terminal period (+1.8 pp) into a \emph{composition effect} (changing income-group weights) and a \emph{within-group effect} (changing mean ranges within each income group). The composition effect is $-0.2$ pp (about $-13\%$ of total change), meaning the shift in the income-group composition of covered country-years would, taken alone, predict lower ranges. The within-group effect is positive: $+2.0$ pp (about $+113\%$ of total change). However, this within-group component is itself a database-proliferation artefact. The mean number of databases covering each country-year has risen substantially over the period, and having more databases per country-year mechanically widens the observed range even if every bilateral MAD is unchanged.

\begin{center}
[Figure~\ref{fig:div_decomp} about here]
\end{center}

Taken together, these results overturn the na\"{i}ve reading of the aggregate trend. The +0.03 Gini points per year in the mean within-country-year range is not driven by existing databases diverging from each other, but by the expansion of the \emph{number} of databases covering each country-year. The aggregate range is widening because the global inequality measurement landscape has become denser, not because its constituent databases are becoming less consistent. This is, in one sense, a reassuring finding about database quality. The long-standing secondary databases have maintained or improved their internal consistency over time. In another sense, it reinforces the central message of this paper. Database proliferation itself is a source of comparability challenges, because each new database brings a new welfare concept, equivalence scale, or estimation methodology that widens the cross-database range for the countries it covers.

\subsection{Pairwise Cross-Database Concordance}\label{subsec:concordance}

Our second approach computes pairwise concordance statistics for all pairs of databases sharing at least twenty overlapping country-year observations, with each database summarised by its median Gini per country-year. We compute the Pearson correlation coefficient (Table~\ref{tab:concordance}, Panel A), the mean absolute difference (MAD) in Gini values, which measures level differences (Table~\ref{tab:concordance}, Panel B), and a measure of change in trend (Table~\ref{tab:concordance}, Panel C).

\begin{center}
[Table~\ref{tab:concordance} about here]
\end{center}

The Pearson correlations in Panel~A, which capture agreement in rank ordering, are uniformly high once each database is summarised by its median Gini per country-year: pairs drawing on common underlying surveys reach 0.98 or above (SEDLAC and WB-PIP at 0.997; ATG and SEDLAC at 0.983), and even the global secondary compilations correlate strongly with the survey-based databases (e.g., SWIID and LIS at 0.975). The lowest correlation involves Eurostat, whose equivalised-income series is the least aligned with the others (0.410 with LIS).

The mean absolute differences (MAD) in Panel~B, which capture agreement in levels, tell a broadly consistent story. Databases sharing primary data sources have very small MADs: SEDLAC and WB-PIP, which rely on the same Latin American household surveys, record a MAD of only 0.07 Gini points, while LIS and OECD, both using income from harmonised surveys for high-income countries, reach 0.88 points. Eurostat is the clear outlier, with MADs ranging from 5.5 to 9.9 points relative to all other databases, a consequence of its exclusive European coverage combined with a specific equivalised income concept; its largest level gap is with LIS (MAD 9.85). The pairs with the highest correlations also have the lowest MADs.

Agreement on levels, moreover, does not guarantee agreement on trends, which are often the object of primary interest for users. Panel~C of Table~\ref{tab:concordance} therefore reports a measure of change in trend. For each pair of databases and each country, we take consecutive years (at most five years apart) in which both databases report a Gini, and compute the share of these matched changes that have the same sign. Most database pairs agree on the direction of change between 60\% and 100\% of the time, with pairs drawing on the same underlying surveys at the top (SEDLAC--WB-PIP 100\%; SEDLAC--UNU-WIDER 90.4\%; LIS--CEPAL 89.6\%). The clear exception is Eurostat, whose directional agreement with most other databases lies between about 49\% and 55\% --- close to a coin flip, the one exception being a higher 71\% agreement with UNU-WIDER --- partly because the year-to-year movements in its equivalised income series are small relative to measurement noise. This remains an important finding. Even where databases agree closely on levels, agreement on short-run \emph{trends} is weaker and concentrated among databases that share surveys, so changes should not be compared, or worse spliced, across databases built on different welfare concepts.

\subsection{The Income--Consumption Gap}\label{subsec:incons}

As documented in previous sections, the choice of money metrics (income or consumption) is a major source of discrepancy across Ginis. We first document the raw income--consumption gaps by region and income group, and then verify that these patterns persist after controlling for the level of development and regional composition.

Table~\ref{tab:inc_cons_region} (Panel A) shows that the mean income--consumption gap is largest in North America (10.2 pp) and Latin America (8.7 pp), and smallest in Europe and Central Asia (2.7 pp). Panel B shows that the gap rises sharply as income level falls, from 3.0 pp for high-income countries to 7.7 pp for low-income countries.

\begin{center}
[Table~\ref{tab:inc_cons_region} about here]
\end{center}

To quantify the income--consumption gap more precisely, controlling for observable heterogeneity in the level of development and regional composition, we estimate the following OLS regression:

\begin{equation}
\Delta G_{ct} = G^{\text{income}}_{ct} - G^{\text{consumption}}_{ct} = \alpha + \beta\,\mathbf{d}^{\text{income group}}_{ct} + \gamma_r + \gamma_t + \varepsilon_{ct}
\label{eq:gap}
\end{equation}

where $\Delta G_{ct}$ is the income--consumption Gini gap for country $c$ in year $t$; $\mathbf{d}^{\text{income group}}$ is a vector of indicator variables for the country's World Bank income group (low, lower-middle, upper-middle) relative to high income; $\gamma_r$ are World Bank region fixed effects; and $\gamma_t$ are decade fixed effects. The sample consists of 1,124 country-year observations for which matched income and consumption Ginis are available.

Table~\ref{tab:inc_cons_regression} presents the regression results, which confirm these patterns after controls. All income-group indicators are positive relative to high income --- low income $+2.9$ (imprecisely estimated), lower-middle income $+3.2$ (significant at 1\%), and upper-middle income $+2.0$ Gini points (significant at 10\%) --- confirming that the income--consumption gap widens as the level of development falls, even after controlling for region and decade effects, although the low-income estimate rests on only 32 matched pairs. Observable heterogeneity in income group and regional composition explains roughly a quarter of the variation in the income--consumption gap ($R^2 = 0.26$), with the remainder reflecting country-specific factors or unobserved heterogeneity in welfare concept definitions.

\begin{center}
[Table~\ref{tab:inc_cons_regression} about here]
\end{center}

\subsection{Welfare Concept, Equivalence Scales, and Temporal Stability}\label{subsec:metareg}

To disentangle the contributions of welfare concept and equivalence scale choice to cross-database Gini variation, while holding constant all country-specific and time-specific factors, we estimate the following equation:

\begin{equation}
G_{ict} = \alpha + \boldsymbol{\beta}_{1}\,\mathbf{d}^{\text{welfare}}_{ict} + \boldsymbol{\beta}_{2}\,\mathbf{d}^{\text{scale}}_{ict} + \mu_c + \mu_t + \varepsilon_{ict}
\label{eq:metareg}
\end{equation}

where $G_{ict}$ is the Gini for observation $i$ (database), country $c$, year $t$; $\mathbf{d}^{\text{welfare}}$ is a vector of welfare-concept indicator variables; $\mathbf{d}^{\text{scale}}$ is a vector of equivalence-scale indicator variables; and $\mu_c$, $\mu_t$ are country and year fixed effects. The identification strategy is within-country-year variation. In other words, we compare Ginis for the same country and year across databases that use different welfare concepts and equivalence scales. The sample consists of 53,523 observations with non-missing welfare concept and equivalence scale information. Standard errors are clustered by country to account for intra-country correlation.

Results in Table~\ref{tab:metareg} (column~1), show that all measurement differences included in the regression contribute to explain variations in the Gini with coefficients that are statistically significant at the 1\% level. Throughout, we interpret these coefficients as descriptive associations between a database's disclosed methodological label and its measured Gini, conditional on country and year, rather than as causal estimates of a welfare-concept effect. Conditional on the fixed effects, the welfare-concept and equivalence-scale indicators yield a within-group $R^2$ of 0.185. In other words, they account for 18.5\% of the Gini variation that remains after absorbing all stable country differences and common year shocks. As we should expect, country and year fixed effects contribute to improve the overall $R^2$ significantly.

The welfare concept associations are all positive relative to the consumption/expenditure reference. Income-based measures are associated with higher Ginis by 3.6 (mixed income concepts), 3.7 (net disposable income), and 5.7 (gross income) percentage points, conditional on country and year fixed effects. The gross-income premium is largest because gross income, before deduction of taxes and social contributions, is substantially more dispersed than net disposable income, which is compressed by progressive taxation and transfers concentrated at the bottom of the distribution. 

The equivalence-scale associations are smaller but statistically significant. The OECD-modified adult-equivalent scale is associated with a Gini reduction of 1.2 points relative to per capita, reflecting the scale's upward adjustment of welfare for small (typically higher-income) households. Household-level unequalised measures are associated with a Gini increase of 2.9 points relative to per capita, consistent with the effective down-weighting of large, lower-income households relative to their population share.

Three important caveats apply when interpreting these results. First, observations from secondary databases are not independent of the primary databases they draw upon (Section~\ref{sec:unified}), so the effective number of independent observations is smaller than the nominal sample size. Clustering standard errors by country mitigates, but does not eliminate, this concern. A fully ``primary-only'' estimation is not feasible. Because each primary database adopts a single welfare concept and scale, the within-country-year contrasts that identify the premia originate almost entirely in the source-documented secondary compilations (WIID, ATG). Column~2, which excludes SWIID (the one secondary database whose values are model-imputed rather than source-recorded) is therefore the most informative robustness check available. It leaves the estimates intact and, if anything, slightly increases most of the estimated premia. Because the identifying variation is concentrated in WIID and ATG, the magnitudes should be read as the average difference between Ginis carrying different concept labels within a country-year, not as the effect of changing the welfare concept while holding the underlying data source fixed. 

Second, the World Inequality Database is excluded from this regression because it does not record welfare-concept or equivalence-scale metadata in machine-readable form. Because WID is the primary driver of divergence at the very top of the Gini distribution (55.5 versus 37--39 for survey-based databases), the regression coefficients should be understood as conservative estimates of the total cross-database divergence attributable to methodological choices. Including WID would widen the estimated income premia substantially. 

Third, our data do not permit us to identify many other important measurement differences such as those in survey design, sampling frames, imputation procedures, top-coding rules, and sub-metric definitions discussed in other parts of the paper. These factors together may well explain an important proportion of the variation of the Gini, and some of these factors may be correlated with country/year, our fixed effects. Unfortunately, metadata provided by database administrators do not provide such level of details, which prevents us from expanding the econometric analysis to these other factors.

A natural follow-up question is whether the welfare-concept and equivalence-scale associations estimated above have remained stable over time, or whether the growing methodological heterogeneity documented in Section~\ref{subsec:variability} has translated into changing coefficient magnitudes. We address this by re-estimating the regression separately for observations before and after 2000 (Table~\ref{tab:metareg}, columns~3--4). The results reveal a clear and substantively important pattern. The income premium over consumption has grown significantly since 2000. In the pre-2000 subsample, net disposable income is associated with a premium of $+1.6$ Gini points relative to consumption (significant at 1\%), while gross income carries $+3.7$ points. In the post-2000 subsample, these premia rise to $+4.2$ and $+6.2$ points respectively, an increase of approximately 2.5 percentage points for both income concepts.

\begin{center}
[Table~\ref{tab:metareg} about here]
\end{center}

An important interpretive question is whether this temporal shift reflects genuine methodological divergence across databases (the same pair of databases agreeing less over time) or merely a change in database composition, as the post-2000 period added many observations from high-income OECD countries where the gross-to-net gap is largest. A direct test is to re-estimate the pre/post-2000 split on a balanced panel of countries. Restricting both subsamples to the 157 countries observed in both periods leaves the estimates essentially unchanged. The net disposable income premium rises from $+1.7$ to $+4.2$ Gini points and the gross income premium from $+3.7$ to $+6.1$, compared with $+1.6$ to $+4.2$ and $+3.7$ to $+6.2$ in the unbalanced split of Table~\ref{tab:metareg} (columns~3--4). The increase in the premia therefore cannot be attributed to the entry of new countries into the sample. It reflects, at least in part, a genuine widening of the gap between income and consumption Ginis for countries covered in both periods, consistent with the secular rise in top-income shares documented for high-income economies since the 1980s. Changes in the composition of databases covering a given country over time may still contribute to the observed pattern, and our data do not allow us to fully disentangle the two mechanisms. Regardless of the precise decomposition, the temporal instability of welfare-concept premia is itself a finding of independent interest. Researchers who pool databases across long time horizons should allow for time-varying welfare-concept adjustments rather than imposing a constant correction.

The equivalisation association shows the opposite pattern. The per capita vs. equivalised difference narrows from $-2.2$ points pre-2000 to $-1.1$ points post-2000. This may reflect the growing prevalence of databases using adult-equivalent scales optimised for specific regional household structures (particularly Latin American databases post-2000), reducing the mechanical compression that adult-equivalent scales impose. The within-group $R^2$ remains stable across periods (0.198--0.199 vs. 0.185 for the full sample), indicating that the explanatory power of welfare concept and scale is period-invariant even though the magnitudes shift.

Finally, we can estimate the importance of the choice between market and disposable income by calculating a redistribution index defined as $R_{ct} = G^{\text{market}}_{ct} - G^{\text{disposable}}_{ct}$ for those databases that provide both market-income (before taxes and transfers) and disposable-income (after taxes and transfers) Ginis. We use the SWIID, which reports standardised market- and disposable-income series for the widest set of countries and years.

\begin{figure}[htbp]
{\centering \includegraphics[width=\linewidth,keepaspectratio]{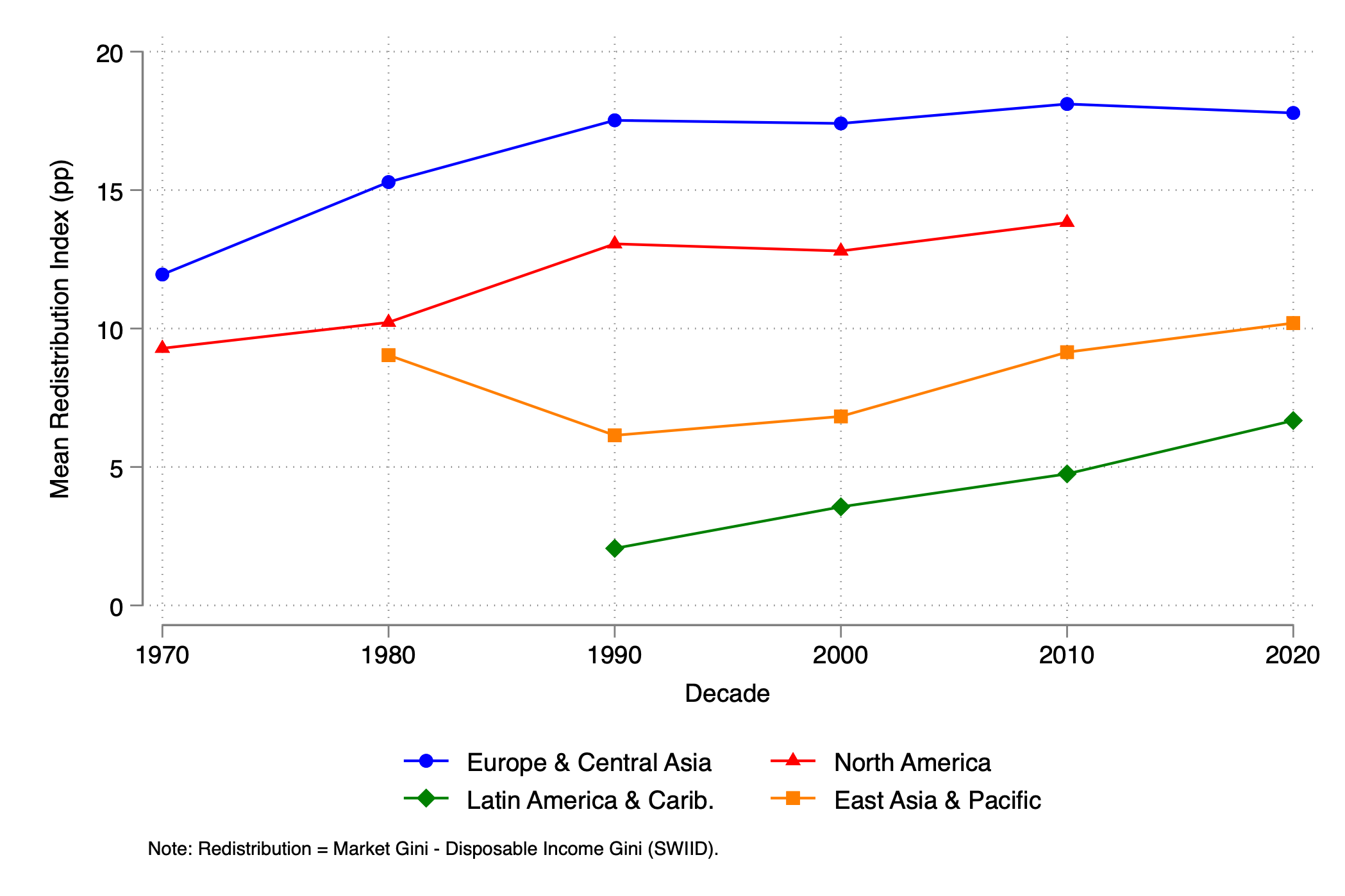}}
\caption{Redistribution Effect (Market vs.\ Disposable Income Gini) Over Time by Region. Decade means are computed over the country-years with matched SWIID market- and disposable-income series; region-decade cells with fewer than 10 country-years are suppressed.}
\label{fig:redistribution}
\end{figure}

Figure~\ref{fig:redistribution} shows trends in the mean redistribution index over time by region where higher values indicate stronger redistribution. For the high-income regions, redistribution is substantial. In Europe and Central Asia, the mean gap between market and disposable Ginis rose markedly through the 1970s and 1980s and has since remained broadly stable at 17--18 Gini points; North America follows a similar profile at a lower level, at around 13 points since the 1990s. The rapid expansion of measured redistribution thus appears to have levelled off after the 1990s, consistent with the slowdown in welfare-state expansion discussed in the retrenchment literature, although the decade means show no outright decline. For the middle-income regions, redistribution is smaller: Latin America and the Caribbean shows a steady rise from about 2 Gini points in the 1990s to almost 7 in the 2020s, while East Asia and the Pacific dipped during the 1990s before recovering to about 10 points in the most recent decades. These patterns confirm that the choice between market and disposable income matters enormously for measured inequality. Studies that compare market incomes across countries will find very different distributional patterns than studies comparing disposable incomes.

\begin{center}
[Figure~\ref{fig:redistribution} about here]
\end{center}

\subsection{Practical Correction Factors}\label{subsec:correction}

The empirical patterns documented in the preceding subsections have direct operational value for researchers who must work with Ginis drawn from different welfare concepts. Three results in particular invite synthesis: the income--consumption gap varies systematically by region and income group (Section~\ref{subsec:incons}), the welfare-concept and equivalence-scale premia identified by the regression are large, statistically significant, and time-varying (Section~\ref{subsec:metareg}), and the gross--net income distinction is quantitatively important for high-income economies with developed fiscal systems. Table~\ref{tab:correction} consolidates these results into a single reference tool, expressing the cross-concept Gini differentials in percentage points so that they can be applied to harmonise estimates across welfare concepts. Because Section~\ref{subsec:metareg} showed these premia to be time-varying, the table should be read as a set of first-order, period-average corrections rather than fixed constants. Where an application spans the pre- and post-2000 periods, the period-specific premia in Table~\ref{tab:metareg} (columns~3--4) are preferable, and the standard errors reported with each factor should be propagated into the adjusted estimates.

Panel~A reports consumption-to-income and net-to-gross adjustments by World Bank region, computed from within-country-year matched pairs. The consumption-to-income gap is small in Europe and Central Asia ($+2.7$ pp), moderate in East Asia and Pacific and the Middle East ($+4.6$ and $+3.4$ pp), and substantially larger in Sub-Saharan Africa ($+7.3$ pp), South Asia ($+7.9$ pp), Latin America ($+8.7$ pp), and North America ($+10.2$ pp). Where labour incomes are more dispersed and self-employment, informal-sector activity, and concentrated capital income are quantitatively important, the welfare-concept choice has greater bearing on measured inequality. The net-to-gross gap behaves differently. It is largest in regions with developed fiscal systems --- North America ($+3.7$ pp), the Middle East ($+3.6$ pp), East Asia and Pacific ($+3.1$ pp), and Europe and Central Asia ($+3.0$ pp) --- and smaller in South Asia, Latin America, and Sub-Saharan Africa, where progressive direct taxation has more limited reach. 

Two distinct concepts should not be confused here. The net-to-gross columns measure the gross--net \emph{welfare-concept} premium, that is, the difference between Ginis computed on income before and after direct taxes and social contributions (both inclusive of transfers). This is conceptually distinct from, and substantially smaller than, the market-to-disposable \emph{redistribution} gap of Section~\ref{subsec:metareg} (Figure~\ref{fig:redistribution}), which additionally strips out transfers; the two adjustments are not interchangeable. These regional adjustments are a first-order correction for researchers who need to combine estimates across welfare concepts within a given region.

Panel~B confirms that the consumption-to-income gap is monotone in the level of development. It rises from $+3.0$ pp for high-income countries to $+5.3$ pp for upper-middle-income, $+6.5$ pp for lower-middle-income, and $+7.7$ pp for low-income economies. The net-to-gross gap exhibits the opposite gradient, narrowing from $+3.2$ pp in high-income countries to $+0.6$ pp in lower-middle-income economies and disappearing entirely in the low-income sample for lack of matched pairs. Together, the two columns of Panel~B sketch a development gradient in which the importance of the welfare-concept choice rises, and the importance of the gross--net distinction falls, as one moves from high-income to low-income countries.

Panel~C provides the regression analogue of the same adjustments, with country and year fixed effects absorbing all stable country-level confounders and common time trends. The estimates are slightly smaller in magnitude than the raw pairwise averages but tell the same story. Net disposable income carries a $+3.7$ pp premium over consumption, gross income a $+5.7$ pp premium, and gross income exceeds net disposable income by $+2.0$ pp. The OECD-modified adult-equivalent scale lowers Ginis by $1.2$ pp relative to per capita, and household-level unequalised measures raise them by $2.9$ pp. Because Panel~C is identified from cross-database variation within country-year cells, these figures are the most directly applicable when the goal is to translate a Gini computed under one methodological convention into the value it would take under another, holding the underlying distribution fixed.

The three panels are best used in combination. A researcher comparing, say, a consumption-based Gini for Bangladesh with a gross-income Gini for the United Kingdom in a cross-country regression would apply the South Asia consumption-to-income correction ($+7.9$ pp) to the Bangladeshi figure and the high-income net-to-gross correction ($+3.2$ pp) to the British figure if the latter is reported on a net basis. A panel study covering Latin America from 1990 to 2020 should also note the post-2000 increase in the gross-income premium documented in Table~\ref{tab:metareg} (columns~3--4). The $+8.7$ pp regional average masks a smaller adjustment in the early period and a larger one in the late period, and period-specific corrections are preferable wherever feasible. 

Four caveats temper the use of the table. First, the Sub-Saharan Africa net-to-gross estimate is based on only fourteen matched pairs and should be treated as indicative. Second, no matched pairs are available for low-income countries on the net-to-gross dimension, so the corresponding cell is left blank. Third, the consumption-to-income corrections are estimated from matched pairs available primarily in SWIID and UNU-WIDER. Their applicability to country-years not well represented in these sources is uncertain. Fourth, and most important, the premia are not stable over time (Section~\ref{subsec:metareg}). The tabulated values are period averages that will degrade as database composition and distributional patterns evolve, so period-specific corrections should be used wherever the sample permits. Subject to these qualifications, the corrections in Table~\ref{tab:correction} provide a transparent, replicable first step toward methodologically consistent inequality comparisons.

\begin{center}
[Table~\ref{tab:correction} about here]
\end{center}

\section{Country Case Studies}\label{sec:casestudies}

The quantitative patterns documented in the previous section gain concreteness when examined at the country level. We present two cases chosen to illustrate the two dominant axes of cross-database discordance identified by the regression: the income--consumption gap, which is the primary source of divergence in middle-income countries, and the gross--net income distinction, which dominates in high-income countries with developed fiscal systems. Colombia and Germany exemplify each axis respectively, and together they demonstrate that comparability problems are not region-specific artefacts but a structural feature of global inequality measurement.

\begin{figure}[htbp]
{\centering \includegraphics[width=\linewidth,keepaspectratio]{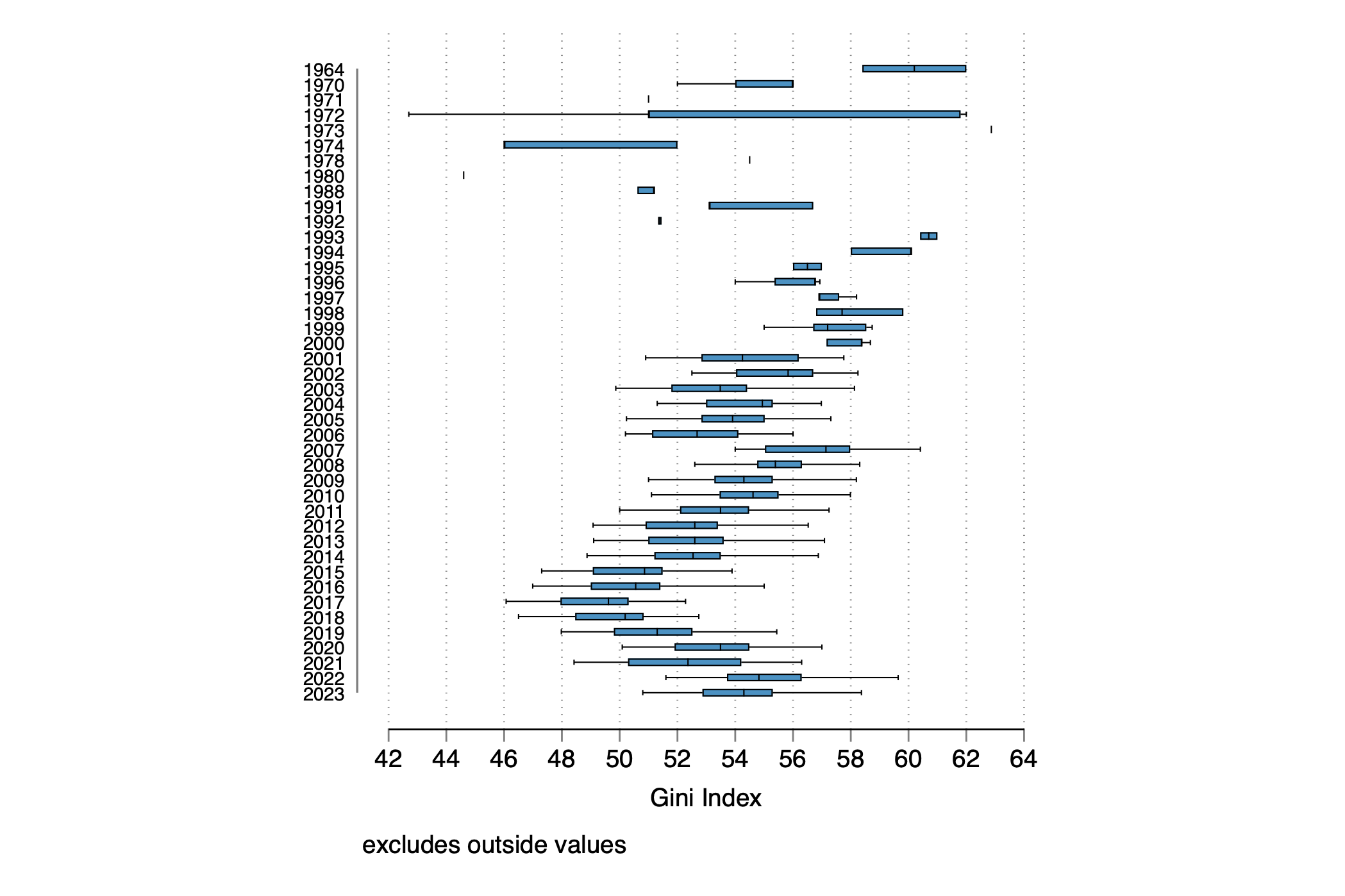}}
\caption{Distribution of Gini Estimates for Colombia by Year (all databases, 1964--2023)}
\label{fig:colombia}
\end{figure}

Colombia is one of the most extensively covered countries in our unified database, with Gini estimates from nine of the thirteen databases spanning 1964 to 2023. Coverage is particularly dense for 2000--2023, when the Latin American regional databases (IDB, SEDLAC, CEPAL), the global secondary databases (WIID, SWIID, ATG), and the global primary databases (WB-PIP, LIS, WID) all report estimates simultaneously. Figure~\ref{fig:colombia} shows boxplots of all Gini estimates for Colombia by year. Several features stand out. The spread of estimates for any given year is large, often exceeding 10 points and reaching 20 or more in some years. The spread is systematically larger in more recent years, when more databases report estimates, rather than reflecting genuine increases in distributional uncertainty. The mean and median Gini track an upward trend through the 1990s and early 2000s, followed by a decline from around 2010, which is broadly consistent with the ``turning point'' documented for Latin American inequality by \citet{LustigEtAl2013}, but the magnitude of the trend depends strongly on which database is used.

\begin{center}
[Figure~\ref{fig:colombia} about here]
\end{center}

\begin{sidewaysfigure}
{\centering \includegraphics[width=\linewidth,keepaspectratio]{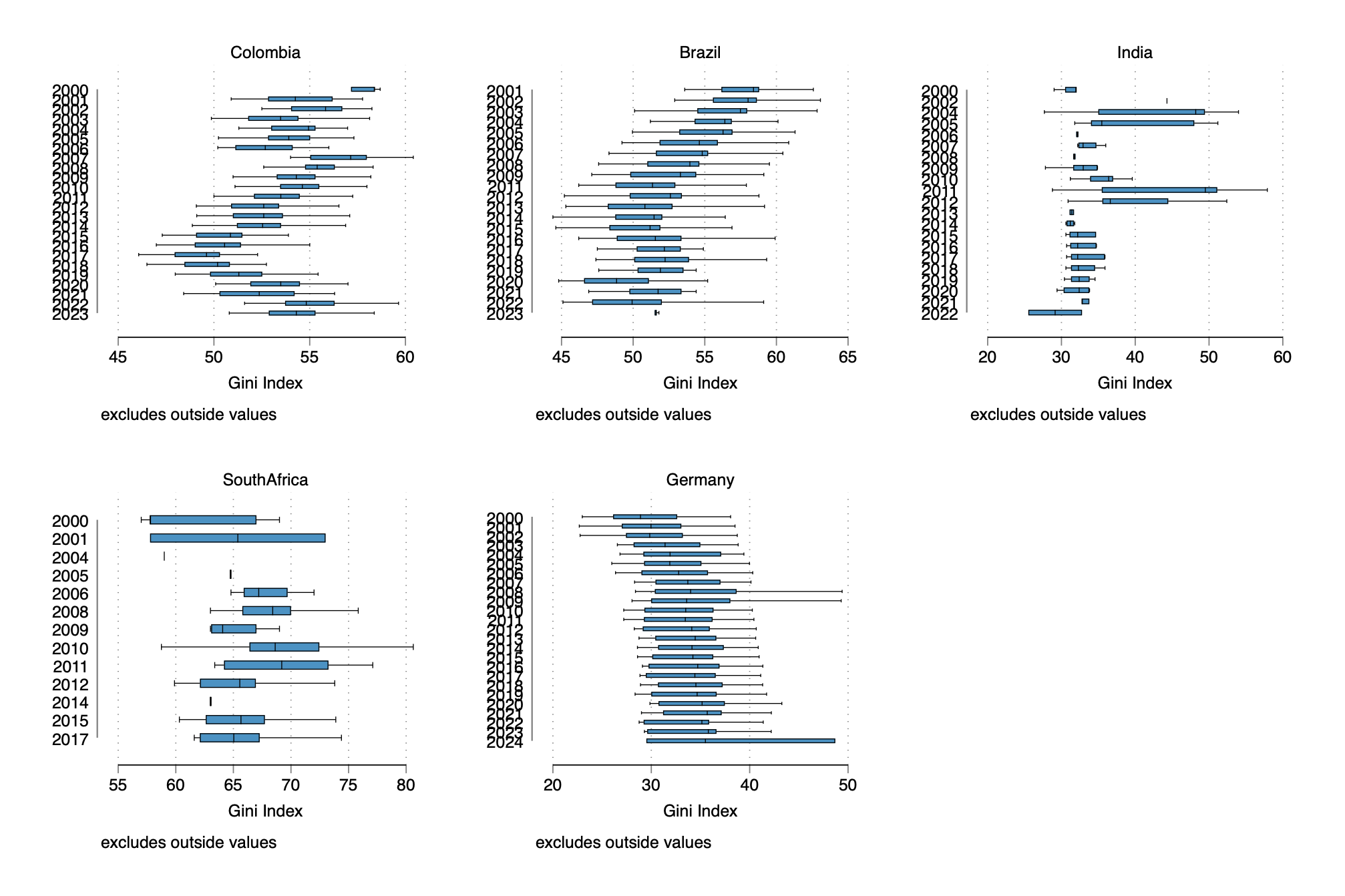}}
\caption{Cross-Database Gini Distributions by Country, 2000--2024 (Colombia, Brazil, India, South Africa, Germany), drawing on all available databases. Boxplots exclude outside values.}
\label{fig:country_compare}
\end{sidewaysfigure}

Figure~\ref{fig:country_compare} shows the resulting cross-database spread for Colombia alongside four other countries. For Colombia the annual range frequently exceeds 10 Gini points. This spread is driven above all by the contrast between the WID estimate (which anchors to national accounts and is typically the highest) and the CEPAL estimate (which uses per-capita household income from national surveys). For most years after 2010 the WID Gini for Colombia is 10--15 points above the CEPAL estimate, the DINA effect in its starkest form.

\begin{center}
[Figure~\ref{fig:country_compare} about here]
\end{center}

\begin{figure}[htbp]
{\centering \includegraphics[width=\linewidth,keepaspectratio]{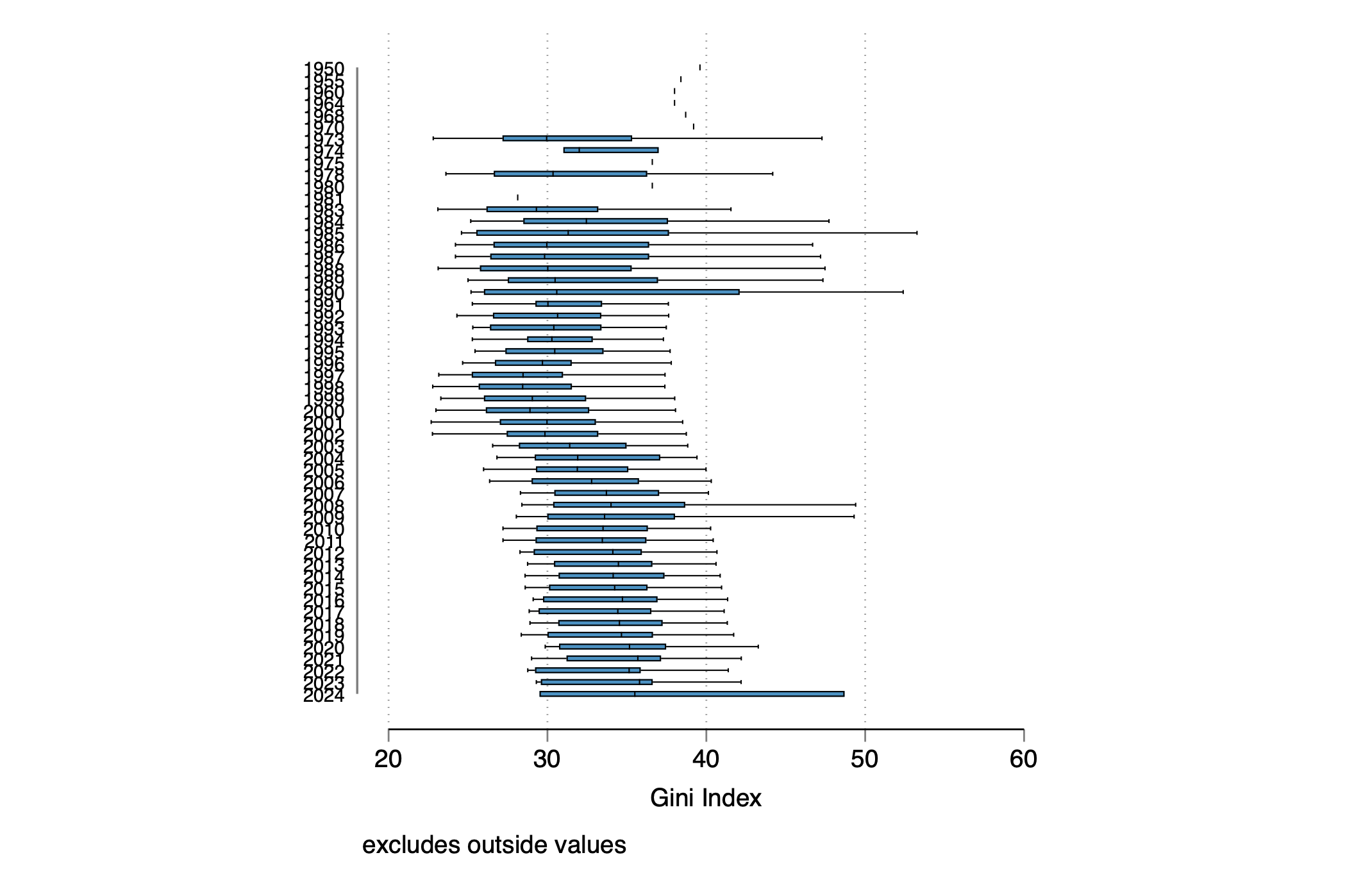}}
\caption{Distribution of Gini Estimates for Germany by Year (all databases, all available years). The gross--net income distinction drives the large cross-database spread; the cross-database interquartile range in 2023 is 7.0 Gini points. Boxplots exclude outside values.}
\label{fig:germany}
\end{figure}

Germany presents a structurally different pattern (Figure~\ref{fig:germany}). It is covered by eight databases (Eurostat, OECD, LIS, WIID, SWIID, ATG, WID, and WB-PIP), but the principal source of divergence is not the income--consumption divide (Germany is measured almost exclusively on an income basis) but the gross--net distinction. Pre-tax income series --- the WID's national accounts-anchored estimates (around 45--49 in recent years) and Eurostat's pre-social-transfer variants --- are dramatically higher than the post-tax equivalised disposable income Ginis reported by Eurostat, LIS, and the OECD (28--31). Germany's well-developed tax and transfer system is associated with a redistribution gap of roughly 20 Gini points between market and disposable income (SWIID standardised series), placing it among the most redistributive economies in the database.

\begin{center}
[Figure~\ref{fig:germany} about here]
\end{center}

Taken together, the two cases carry a clear methodological lesson. In Colombia, the key fault line runs between databases measuring welfare as household income and those measuring it as consumption or anchoring to national accounts. A researcher drawing on different databases to extend a time series or fill gaps would risk conflating genuine distributional change with a measurement convention shift. In Germany, the fault line runs between databases recording market income and those recording post-redistribution disposable income, naively pooling the two would attribute to Germany a level of inequality comparable to highly unequal developing economies, whereas the fiscal system in fact compresses the net distribution substantially. Both cases confirm that cross-database comparability problems are not marginal concerns but affect the sign and magnitude of measured inequality in economically consequential ways.

\section{Conclusion and Recommendations}\label{sec:conclusion}

The global effort to measure income inequality has generated an unprecedented wealth of data over the past three decades, yet the proliferation of databases has not resolved the fundamental challenge of comparability. The thirteen databases brought together here differ in welfare concept, reference unit, sub-metric construction, post-survey adjustment, and coverage, and these differences produce large and systematic discrepancies in measured inequality.

Three findings stand out. First, within-country-year Gini ranges across databases are large in level and have been widening in the aggregate, but a balanced-pair analysis and shift-share decomposition show that the widening is driven by database proliferation rather than by long-standing databases drifting apart from each other. Inter-database consistency among the major long-running sources has actually been stable or improving. Second, welfare-concept and equivalence-scale choices account for a substantial share of cross-database disagreement, with income-based measures systematically yielding higher Ginis than consumption-based ones. This premium has grown markedly since 2000, so correction factors calibrated on historical data cannot be applied unchanged to contemporary series. Third, the practical correction factors consolidated in Table~\ref{tab:correction} provide a transparent, replicable first step toward methodologically consistent inequality comparisons, with regional and income-group differentials that mirror the geography of welfare measurement and redistributive capacity.

The findings consolidate into a short set of operational rules for anyone using cross-country Gini data.

\begin{enumerate}
\item \textbf{Treat database choice as a research decision.} State which database and which vintage (download date, version, DOI where available) underlie the analysis. Revisions can alter historical series and break replicability.

\item \textbf{Prefer a single welfare concept.} Where feasible, restrict the sample to one welfare concept and one equivalence scale. This forgoes coverage but removes the largest source of cross-database divergence.

\item \textbf{If mixing is unavoidable, correct explicitly.} Control for welfare concept and equivalence scale, or apply the correction factors of Table~\ref{tab:correction}, using the region- or income-group-specific values and allowing for the post-2000 increase in the income premia (Table~\ref{tab:metareg}, columns~3--4), treating the tabulated factors as first-order corrections and propagating their standard errors into the adjusted figures. Do not confuse the gross--net tax correction with the much larger market--disposable redistribution gap.

\item \textbf{Never splice WID/DINA series with survey-based series.} National accounts-anchored estimates are systematically higher (mean 55.5 versus 37--39) and follow different revision dynamics. Combining them without adjustment conflates methodology with distributional change.

\item \textbf{Be especially cautious with trends.} Direction-of-change agreement across databases is far from perfect (Table~\ref{tab:concordance}, Panel~C). Even databases that agree closely on levels can disagree on the sign of short-run changes. Trend analysis should rely on a single database per country, and structural breaks from survey redesigns should not be read as distributional change.

\item \textbf{Demand and reward metadata disclosure.} Every Gini observation should carry machine-readable indicators of welfare concept, income type, equivalence scale, reference period, sharing unit, geographic and population coverage, and the treatment of top and bottom incomes. Versioned, citable releases with stable identifiers should become the norm.
\end{enumerate}

The analysis is not without limitations. The regression identifies only two of the many possible sources of discordance and yields conditional associations rather than causal parameters. The World Inequality Database is excluded because it does not record welfare-concept metadata, which makes our estimates conservative, and the temporal instability of welfare-concept premia implies that fixed correction factors will degrade as database composition and distributional patterns evolve. The road to truly comparable global inequality data therefore remains long, but the centrality of inequality to social welfare, economic development, and public policy makes the investment worthwhile. The world's Ginis deserve the same transparency and observational standards that international statistical norms now demand for macroeconomic aggregates.


\paragraph{Data and replication.} The unified dataset, harmonisation code, and the replication scripts for all tables and figures are submitted with the manuscript and will be deposited in a public repository upon acceptance. All thirteen source databases are publicly accessible at the URLs cited in the References; the specific vintages used in this analysis, the harmonised variable coding, and the replication instructions are documented in the accompanying Online Supplement.


\bibliography{References}


\clearpage
\begin{table}[p]
    \centering
    \small
    \caption{Prior Cross-Database Assessments of Inequality Data Comparability\label{tab:priorstudies}}
    \adjustbox{max size={\textwidth}{0.82\textheight}}{%
    \begin{tabular}{p{3.6cm}p{3.4cm}p{2.4cm}p{5.2cm}}
    \toprule
    Study & Data compared & Scope & Main message \\
    \midrule
    \citet{AtkinsonMicklewright1992} & National sources & Eastern Europe & Comparison requires source-level documentation of definitions \\
    \citet{GottschalkSmeeding1997} & LIS microdata vs national sources & OECD & Harmonised microdata overturn rankings based on unadjusted sources \\
    \citet{AtkinsonBrandolini2001} & Deininger--Squire, OECD sources & OECD & Secondary datasets can mislead; welfare-concept dummies are insufficient \\
    \citet{Deaton2005} & Surveys vs national accounts & Global & Survey and national-accounts means diverge systematically \\
    \citet{Jenkins2015}; \citet{Solt2015} & WIID vs SWIID & Global & Trade-off between source documentation (WIID) and imputed comparability (SWIID) \\
    \citet{FerreiraEtAl2015} & Cross-national inequality databases & Global & Different databases give different pictures of trends; call for systematic assessment \\
    \citet{PikettySaezZucman2018}; \citet{AutenSplinter2024} & US tax data (DINA) & United States & Same administrative data, divergent trends under different methodological choices \\
    \citet{Lustig2019} & Top-income corrections & Global & ``Missing rich'' biases levels and trends; corrections are method-sensitive \\
    \textbf{This paper} & 13 global and regional databases & Global, 1867--2024 & Quantified welfare-concept and scale premia; correction factors; divergence driven by database proliferation \\
    \bottomrule
    \end{tabular}}
\end{table}

\clearpage
\begin{sidewaystable}[p]
\centering
\footnotesize
\setlength{\tabcolsep}{4pt}
\renewcommand{\arraystretch}{0.9}
\caption{Overview of Databases Included in the Unified Collection\label{tab:databases}}
\adjustbox{max size={0.92\textheight}{0.85\textwidth}}{%
\begin{tabular}{lcrclcp{1.3cm}cccl}
\toprule
 &&   & & & Wellbeing &  &  &  \\
Dataset & Primary & \# Obs & Countries & Time Range & Concept & Level & Sharing & Reference & Gini & Source\\
 & Dataset & & & & (\% Cons.)  & Unit & Unit &  & \\
\hline
\hspace{1em}GLOBAL &  &  &  &  &  & &  & & &\\
\hline
\hspace{1em}ATG & no & 5,121 & 175 & 1948--2017 & 35.79 & overall &HH & Mixed$^{\dagger}$& 38.76 & \citet{Milanovic_2019}\\
\hspace{1em}LIS & yes & 930 & 52 & 1963--2023 & 0.00 & overall & HH& AE& 33.70 &\citet{LIS_2024}\\
\hspace{1em}OECD & no & 616 & 45 & 1976--2023 & 0.00 & overall  & HH& AE&  31.85 &\citet{OECD_2024}\\
\hspace{1em}UNU-WIDER & no & 26,161 & 200 & 1867--2023 & 11.77 &overall & HH& Mixed$^{*}$& 37.06 &\citet{UNUWIDER_2025}\\
\hspace{1em}WB-PIP & yes & 2,504 & 172 & 1963--2024 & 38.70  & overall& HH& PC& 37.17 &\citet{PIP_2024}\\
\hspace{1em}WID & yes & 10,419 & 217 & 1900--2023 & 0.00  &overall & Tax Unit& PC& 55.45 &\citet{BlanchetEtAl2024}\\
\hspace{1em}SWIID & no & 26,900 & 199 & 1960--2023 & 13.97  & overall& Ind & Mixed$^{**}$& 39.49 & \citet{Solt_2020}\\
&  &  &  &  &  &  & HH  & & & \\
&\\
\hline
\hspace{1em}REGIONAL &  &  &  &  &  & &  & & & \\
\hline
\hspace{1em}Africa &  &  &  &  &  & &  & & & \\
\hspace{1em}Afristat &  no & 84 & 20 & 1994--2023 & ? & overall& ?& ?&  39.40 &  \citet{Afristat_2025}\\
\hspace{1em}East Asia and the Pacific &  &  &  &  &  & &  & & &\\
\hspace{1em}ADB &  no & 322 & 40 & 2000--2023 & 0.00  & overall&HH & PC&  34.52 &\citet{ADB_2024, ADB_2025}\\
\hspace{1em}Europe &  &  &  &  &  & &  & & & \\
\hspace{1em}Eurostat &  yes & 2,102 & 37 & 2003--2024 & 0.00  & overall& HH& AE&  37.71$^{\S}$ & \citet{Eurostat_2024}$^{\P}$\\
&  &  &  &  &  & age $<$ 18 &   & & & \\
\hspace{1em}Latin America and the Caribbean &  &  &  &  &  & &  & & & \\
\hspace{1em}CEPAL &  yes & 887 & 18 & 2000--2023 & 0.00  & overall& HH& PC&  47.10 & \citet{CEPAL_2024}\\
&  &  &  &  &  & urban/rural &   & & & \\
\hspace{1em}IDB &  yes & 45,656 & 26 & 1970--2023 & 0.00  & overall& Ind& PC&  36.69 & \citet{IADB_2025}\\
&  &  &  &  &  & urban/rural & HH  & & & \\
&  &  &  &  &  & gender &   & & & \\
&  &  &  &  &  & migration &   & & & \\
&  &  &  &  &  & ethnicity &   & & & \\
&  &  &  &  &  & disability &   & & & \\
&  &  &  &  &  & quintiles &   & & & \\
\hspace{1em}SEDLAC &  yes & 649 & 23 & 2000--2024 & 1.08  & overall& HH& PC/AE&  48.22& \citet{SEDLAC_2024}\\
&  &  &  &  &  & urban &   & & & \\
&  &  &  &  &  & main &   & & & \\
&  &  &  &  &  & cities&   & & & \\
\hline
\hspace{1em}Total & &  122,351 & 222$^{\ddagger}$ & 1867--2024 & 7.88  & overall& HH& &  39.19 &\\
\bottomrule
\multicolumn{10}{l}{$^*$ 39\% PC; 41\% AE; 20\% Household Based}\\
\multicolumn{10}{l}{$^\dagger$ 72\% PC; 8\% Household Based}\\
\multicolumn{10}{l}{$^{**}$ 32\% PC; 52\% AE; 15\% Household Based}\\
\multicolumn{10}{l}{$^{\ddagger}$ Distinct countries and territories across all databases.}\\
\multicolumn{10}{l}{$^{\P}$ Eurostat series tessi190, tessi191, ilc\_di12b, and ilc\_di12c.}\\
\multicolumn{10}{l}{``?'' = information not documented by the source database. PC = per capita; AE = adult equivalent; HH = household; Ind = individual.}\\
\multicolumn{10}{l}{$^{\S}$ Pools Eurostat's three income concepts; the disposable equivalised income series alone averages $\approx$30, comparable to LIS (33.7) and OECD (31.9).}\\
\end{tabular}}
\end{sidewaystable}


\clearpage
\begin{table}[p]
    \centering
    \small
    \caption{Database Coverage by World Bank Income Group (unique country-year observations)\label{tab:coverage_income}}
    \adjustbox{max size={\textwidth}{0.82\textheight}}{%
    \begin{tabular}{lrrrrr}
    \toprule
    Dataset & Low & Lower-Mid & Upper-Mid & High & Total \\
    \midrule
    SWIID       & 181 & 614  & 1,009 & 1,763 & 3,567 \\
    UNU-WIDER   & 160 & 591  & 948   & 1,751 & 3,450 \\
    WB-PIP      & 123 & 419  & 729   & 1,075 & 2,346 \\
    ATG         & 131 & 464  & 696   & 1,050 & 2,341 \\
    LIS         & 7   & 11   & 176   & 736   & 930   \\
    Eurostat    & 0   & 0    & 76    & 602   & 678   \\
    OECD        & 0   & 2    & 61    & 553   & 616   \\
    IDB         & 1   & 64   & 214   & 109   & 388   \\
    SEDLAC      & 0   & 68   & 200   & 56    & 324   \\
    ADB         & 5   & 109  & 152   & 56    & 322   \\
    CEPAL       & 0   & 60   & 173   & 63    & 296   \\
    Afristat    & 49  & 34   & 1     & 0     & 84    \\
    \bottomrule
    \end{tabular}}
    \par\smallskip\noindent\footnotesize\textit{Note:} Counts unique country-year observations per database, excluding WID due to its comprehensive but structurally different coverage.
\end{table}

\clearpage
\begin{table}[p]
    \centering
    \small
    \caption{Cross-Database Gini Variability by Region and Income Group (country-years with $\geq 2$ database observations)\label{tab:variability_region}}
    \adjustbox{max size={\textwidth}{0.82\textheight}}{%
    \begin{tabular}{lrrrrr}
    \toprule
    & \# Obs & Mean Range & Mean SD & Median Range & Max Range \\
    \midrule
    \multicolumn{6}{l}{\textit{Panel A: By World Bank Region}} \\
    East Asia \& Pacific (EAS)          & 406   & 2.60 & 1.39 & 2.00 & 22.00 \\
    Europe \& Central Asia (ECS)        & 1,532 & 5.26 & 2.21 & 4.21 & 23.10 \\
    Latin America \& Caribbean (LCN)    & 631   & 4.18 & 1.84 & 3.31 & 43.42 \\
    Middle East \& N. Africa (MEA)      & 200   & 2.36 & 1.24 & 1.42 & 18.63 \\
    North America (NAC)                 & 127   & 6.40 & 2.64 & 6.45 & 10.60 \\
    South Asia (SAS)                    & 182   & 2.79 & 1.46 & 1.76 & 21.32 \\
    Sub-Saharan Africa (SSF)            & 341   & 2.93 & 1.60 & 1.50 & 25.50 \\
    \textbf{Total}                      & \textbf{3,419} & \textbf{4.25} & \textbf{1.90} & \textbf{3.27} & \textbf{43.42} \\
    \midrule
    \multicolumn{6}{l}{\textit{Panel B: By World Bank Income Group}} \\
    High income         & 1,668 & 5.46 & 2.34 & 4.53 & 43.42 \\
    Upper middle income & 969   & 3.44 & 1.55 & 2.65 & 31.55 \\
    Lower middle income & 600   & 2.80 & 1.45 & 1.89 & 25.50 \\
    Low income          & 182   & 2.37 & 1.26 & 1.18 & 20.78 \\
    \textbf{Total}      & \textbf{3,419} & \textbf{4.25} & \textbf{1.90} & \textbf{3.27} & \textbf{43.42} \\
    \bottomrule
    \end{tabular}}
    \par\smallskip\noindent\footnotesize\textit{Note:} Each database enters at its median Gini per country-year (so that databases reporting several welfare concepts or sources contribute a single value). Range = maximum minus minimum Gini across databases for the same country-year (Gini points). SD = cross-database standard deviation.
\end{table}

\clearpage
\begin{sidewaystable}[p]
\centering
\small
\caption{Pairwise Pearson Correlation and Mean Absolute Difference (pp) Across Databases\label{tab:concordance}}

\setlength{\tabcolsep}{3pt}
\renewcommand{\arraystretch}{0.95}

\adjustbox{max size={\textheight}{0.92\textwidth}}{%
\begin{tabular}{lrrrrrrrrrr}
\toprule
 & ATG & CEPAL & Eurostat & IDB & LIS & OECD & SEDLAC & SWIID & UNU-WIDER & WB-PIP \\
\midrule
\multicolumn{11}{l}{\textit{Panel A: Pearson Correlation}} \\
ATG       & 1.000 & 0.937 & 0.534 & 0.826 & 0.948 & 0.926 & 0.983 & 0.947 & 0.950 & 0.977 \\
CEPAL     & 0.937 & 1.000 & .     & 0.830 & 0.862 & 0.492 & 0.964 & 0.951 & 0.954 & 0.935 \\
Eurostat  & 0.534 & .     & 1.000 & .     & 0.410 & 0.614 & .     & 0.603 & 0.574 & 0.537 \\
IDB       & 0.826 & 0.830 & .     & 1.000 & 0.857 & 0.550 & 0.835 & 0.643 & 0.720 & 0.821 \\
LIS       & 0.948 & 0.862 & 0.410 & 0.857 & 1.000 & 0.963 & 0.822 & 0.975 & 0.966 & 0.958 \\
OECD      & 0.926 & 0.492 & 0.614 & 0.550 & 0.963 & 1.000 & 0.414 & 0.943 & 0.952 & 0.929 \\
SEDLAC    & 0.983 & 0.964 & .     & 0.835 & 0.822 & 0.414 & 1.000 & 0.970 & 0.982 & 0.997 \\
SWIID     & 0.947 & 0.951 & 0.603 & 0.643 & 0.975 & 0.943 & 0.970 & 1.000 & 0.954 & 0.968 \\
UNU-WIDER & 0.950 & 0.954 & 0.574 & 0.720 & 0.966 & 0.952 & 0.982 & 0.954 & 1.000 & 0.968 \\
WB-PIP    & 0.977 & 0.935 & 0.537 & 0.821 & 0.958 & 0.929 & 0.997 & 0.968 & 0.968 & 1.000 \\
\midrule
\multicolumn{11}{l}{\textit{Panel B: Mean Absolute Difference (Gini points)}} \\
ATG       & 0.00 & 1.31 & 9.15 & 2.82 & 3.28 & 2.04 & 0.55 & 1.81 & 1.79 & 0.86 \\
CEPAL     & 1.31 & 0.00 & .     & 2.85 & 3.34 & 2.77 & 1.20 & 1.23 & 1.28 & 1.35 \\
Eurostat  & 9.15 & .    & 0.00  & .    & 9.85 & 7.83 & .    & 5.86 & 5.47 & 7.11 \\
IDB       & 2.82 & 2.85 & .     & 0.00 & 5.55 & 5.69 & 3.22 & 4.28 & 4.25 & 3.33 \\
LIS       & 3.28 & 3.34 & 9.85 & 5.55 & 0.00 & 0.88 & 3.21 & 3.63 & 2.90 & 2.58 \\
OECD      & 2.04 & 2.77 & 7.83 & 5.69 & 0.88 & 0.00 & 2.10 & 2.81 & 2.73 & 1.80 \\
SEDLAC    & 0.55 & 1.20 & .     & 3.22 & 3.21 & 2.10 & 0.00 & 0.82 & 0.54 & 0.07 \\
SWIID     & 1.81 & 1.23 & 5.86  & 4.28 & 3.63 & 2.81 & 0.82 & 0.00 & 1.54 & 1.43 \\
UNU-WIDER & 1.79 & 1.28 & 5.47  & 4.25 & 2.90 & 2.73 & 0.54 & 1.54 & 0.00 & 1.38 \\
WB-PIP    & 0.86 & 1.35 & 7.11  & 3.33 & 2.58 & 1.80 & 0.07 & 1.43 & 1.38 & 0.00 \\
\midrule
\multicolumn{11}{l}{\textit{Panel C: Direction-of-Change Agreement (\% of matched consecutive changes with same sign)}} \\
ATG       & ---  & 83.2 & 50.8 & 79.1 & 74.0 & 78.1 & 91.2 & 74.1 & 71.0 & 85.4 \\
CEPAL     & 83.2 & ---  & .    & 81.8 & 89.6 & 77.8 & 86.8 & 86.2 & 87.5 & 86.4 \\
Eurostat  & 50.8 & .    & ---  & .    & 55.2 & 48.8 & .    & 49.4 & 70.9 & 49.2 \\
IDB       & 79.1 & 81.8 & .    & ---  & 85.0 & 69.0 & 84.3 & 76.7 & 79.0 & 81.7 \\
LIS       & 74.0 & 89.6 & 55.2 & 85.0 & ---  & 83.6 & 86.0 & 83.4 & 65.2 & 84.6 \\
OECD      & 78.1 & 77.8 & 48.8 & 69.0 & 83.6 & ---  & 75.0 & 73.7 & 60.3 & 81.2 \\
SEDLAC    & 91.2 & 86.8 & .    & 84.3 & 86.0 & 75.0 & ---  & 86.9 & 90.4 & 100.0 \\
SWIID     & 74.1 & 86.2 & 49.4 & 76.7 & 83.4 & 73.7 & 86.9 & ---  & 69.7 & 81.6 \\
UNU-WIDER & 71.0 & 87.5 & 70.9 & 79.0 & 65.2 & 60.3 & 90.4 & 69.7 & ---  & 71.8 \\
WB-PIP    & 85.4 & 86.4 & 49.2 & 81.7 & 84.6 & 81.2 & 100.0 & 81.6 & 71.8 & ---  \\
\bottomrule
\end{tabular}}
\par\smallskip\noindent\footnotesize\textit{Note:} Pairs with fewer than 20 overlapping country-year observations (Panels A--B) or fewer than 20 matched consecutive changes (Panel C) are denoted by a period. Panel C: for each country, consecutive years at most five years apart in which both databases report a Gini; entries give the percentage of such changes with the same sign in both databases.
\end{sidewaystable}

\clearpage
\begin{table}[p]
    \centering
    \small
    \caption{Income--Consumption Gini Gap by Region and Income Group\label{tab:inc_cons_region}}
    \adjustbox{max size={\textwidth}{0.82\textheight}}{%
\begin{tabular}{lllll}
\cline{1-5}
\multicolumn{1}{c}{} &
  \multicolumn{1}{r}{\# Obs} &
  \multicolumn{1}{r}{Mean Gap} &
  \multicolumn{1}{r}{Median Gap} &
  \multicolumn{1}{r}{P75} \\
\cline{1-5}
\multicolumn{1}{l}{\textit{Panel A: By World Bank Region}} &
  \multicolumn{1}{r}{} &
  \multicolumn{1}{r}{} &
  \multicolumn{1}{r}{} &
  \multicolumn{1}{r}{} \\
\multicolumn{1}{l}{\hspace{3em}East Asia \& Pacific} &
  \multicolumn{1}{r}{152} &
  \multicolumn{1}{r}{4.58} &
  \multicolumn{1}{r}{4.28} &
  \multicolumn{1}{r}{7.42} \\
\multicolumn{1}{l}{\hspace{3em}Europe \& Central Asia } &
  \multicolumn{1}{r}{544} &
  \multicolumn{1}{r}{2.69} &
  \multicolumn{1}{r}{2.23} &
  \multicolumn{1}{r}{5.75} \\
\multicolumn{1}{l}{\hspace{3em}Latin America \& Caribbean} &
  \multicolumn{1}{r}{118} &
  \multicolumn{1}{r}{8.67} &
  \multicolumn{1}{r}{7.91} &
  \multicolumn{1}{r}{12.23} \\
\multicolumn{1}{l}{\hspace{3em}Middle East \& N. Africa} &
  \multicolumn{1}{r}{96} &
  \multicolumn{1}{r}{3.35} &
  \multicolumn{1}{r}{3.10} &
  \multicolumn{1}{r}{5.17} \\
\multicolumn{1}{l}{\hspace{3em}North America} &
  \multicolumn{1}{r}{51} &
  \multicolumn{1}{r}{10.21} &
  \multicolumn{1}{r}{11.26} &
  \multicolumn{1}{r}{12.75} \\
\multicolumn{1}{l}{\hspace{3em}South Asia} &
  \multicolumn{1}{r}{76} &
  \multicolumn{1}{r}{7.93} &
  \multicolumn{1}{r}{6.97} &
  \multicolumn{1}{r}{11.76} \\
\multicolumn{1}{l}{\hspace{3em}Sub-Saharan Africa} &
  \multicolumn{1}{r}{87} &
  \multicolumn{1}{r}{7.30} &
  \multicolumn{1}{r}{6.29} &
  \multicolumn{1}{r}{12.93} \\
\multicolumn{1}{l}{\hspace{3em}\textbf{Total}} &
  \multicolumn{1}{r}{1,124} &
  \multicolumn{1}{r}{4.68} &
  \multicolumn{1}{r}{4.31} &
  \multicolumn{1}{r}{7.52} \\
\cline{1-5}
\multicolumn{1}{l}{\hspace{1em}\textit{Panel B: By World Bank Income Group}} &
  \multicolumn{1}{r}{} &
  \multicolumn{1}{r}{} &
  \multicolumn{1}{r}{} &
  \multicolumn{1}{r}{} \\
\multicolumn{1}{l}{\hspace{3em}High income} &
  \multicolumn{1}{r}{446} &
  \multicolumn{1}{r}{2.96} &
  \multicolumn{1}{r}{2.39} &
  \multicolumn{1}{r}{5.75} \\
\multicolumn{1}{l}{\hspace{3em}Upper middle income} &
  \multicolumn{1}{r}{412} &
  \multicolumn{1}{r}{5.27} &
  \multicolumn{1}{r}{4.92} &
  \multicolumn{1}{r}{8.29} \\
\multicolumn{1}{l}{\hspace{3em}Lower middle income} &
  \multicolumn{1}{r}{234} &
  \multicolumn{1}{r}{6.52} &
  \multicolumn{1}{r}{5.83} &
  \multicolumn{1}{r}{10.11} \\
\multicolumn{1}{l}{\hspace{3em}Low income} &
  \multicolumn{1}{r}{32} &
  \multicolumn{1}{r}{7.71} &
  \multicolumn{1}{r}{6.76} &
  \multicolumn{1}{r}{12.93} \\
\multicolumn{1}{l}{\hspace{3em}\textbf{Total}} &
  \multicolumn{1}{r}{1,124} &
  \multicolumn{1}{r}{4.68} &
  \multicolumn{1}{r}{4.31} &
  \multicolumn{1}{r}{7.52} \\
\cline{1-5}
\end{tabular}}
\end{table}

\clearpage
\begin{table}[p]
    \centering
    \small
    \caption{OLS Regression: Determinants of the Income--Consumption Gini Gap\label{tab:inc_cons_regression}}
    \adjustbox{max size={\textwidth}{0.82\textheight}}{%
    \begin{tabular}{lc}
    \toprule
    Variable & (1) Gap (pp) \\
    \midrule
    \multicolumn{2}{l}{\textit{World Bank income group (ref: high income)}} \\
    \quad Low income & $2.938$ \\
     & $(2.107)$ \\
    \quad Lower middle income & $3.168^{***}$ \\
     & $(0.922)$ \\
    \quad Upper middle income & $1.993^{*}$ \\
     & $(1.066)$ \\
    \midrule
    \multicolumn{2}{l}{\textit{Decade fixed effects (ref: 2000s)}} \\
    \quad 1950 & $-7.458^{***}$ \\
     & $(1.816)$ \\
    \quad 1960 & $1.258$ \\
     & $(1.355)$ \\
    \quad 1970 & $-1.369$ \\
     & $(1.102)$ \\
    \quad 1980 & $0.229$ \\
     & $(0.649)$ \\
    \quad 1990 & $0.295$ \\
     & $(0.539)$ \\
    \quad 2010 & $1.276^{**}$ \\
     & $(0.513)$ \\
    \quad 2020 & $1.400^{*}$ \\
     & $(0.743)$ \\
    \midrule
    \multicolumn{2}{l}{\textit{Region fixed effects (ref: Sub-Saharan Africa)}} \\
    \quad East Asia \& Pacific & $-1.537$ \\
     & $(1.503)$ \\
    \quad Europe \& Central Asia & $-2.997^{**}$ \\
     & $(1.232)$ \\
    \quad Latin America \& Caribbean & $2.114$ \\
     & $(2.177)$ \\
    \quad Middle East \& N.\ Africa & $-3.008^{**}$ \\
     & $(1.195)$ \\
    \quad North America & $5.769^{***}$ \\
     & $(1.959)$ \\
    \quad South Asia & $1.273$ \\
     & $(1.514)$ \\
    \midrule
    Constant & $4.123^{***}$ \\
     & $(1.282)$ \\
    \midrule
    Observations & 1,124 \\
    $R^2$ & 0.261 \\
    \bottomrule
    \end{tabular}}
    \par\smallskip\noindent\footnotesize\textit{Notes:} Standard errors clustered by country in parentheses. $^{***}\,p<0.01$; $^{**}\,p<0.05$; $^{*}\,p<0.10$. Omitted categories: high income; decade 2000s; region Sub-Saharan Africa.
\end{table}

\clearpage
\begin{table}[p]
    \centering
    \small
    \caption{Regression: Association of Welfare Concept and Equivalence Scale with the Gini --- Robustness and Period Heterogeneity\label{tab:metareg}}
    \adjustbox{max size={\textwidth}{0.82\textheight}}{%
    \begin{tabular}{lcccc}
    \toprule
    & (1) Full sample & (2) Excl.\ SWIID & (3) Pre-2000 & (4) Post-2000 \\
    Variable & Gini (pp) & Gini (pp) & Gini (pp) & Gini (pp) \\
    \midrule
    \multicolumn{5}{l}{\textit{Welfare concept (ref: consumption/expenditure)}} \\
    \quad Net disposable income & $3.698^{***}$ & $4.512^{***}$ & $1.647^{***}$ & $4.239^{***}$ \\
                              & $(0.539)$ & $(0.567)$ & $(0.601)$ & $(0.629)$ \\
    \quad Mixed income concepts & $3.605^{***}$ & $3.411^{***}$ & $2.208^{***}$ & $3.916^{***}$ \\
                              & $(0.556)$ & $(0.577)$ & $(0.593)$ & $(0.663)$ \\
    \quad Gross income         & $5.736^{***}$ & $6.143^{***}$ & $3.704^{***}$ & $6.164^{***}$ \\
                              & $(0.558)$ & $(0.596)$ & $(0.603)$ & $(0.661)$ \\
    \midrule
    \multicolumn{5}{l}{\textit{Equivalence scale (ref: per capita)}} \\
    \quad OECD-modified AE scale & $-1.203^{***}$ & $-1.305^{***}$ & $-2.210^{***}$ & $-1.145^{***}$ \\
                              & $(0.178)$ & $(0.183)$ & $(0.212)$ & $(0.183)$ \\
    \quad Household (unequalised) & $2.852^{***}$ & $2.997^{***}$ & $2.054^{***}$ & $2.843^{***}$ \\
                              & $(0.384)$ & $(0.370)$ & $(0.480)$ & $(0.428)$ \\
    \midrule
    Observations & 53,523 & 32,885 & 13,364 & 40,159 \\
    Countries    & 201    & 196    & 163    & 195    \\
    Within-group $R^2$ & $0.185$ & $0.166$ & $0.198$ & $0.199$ \\
    Country FE & Yes & Yes & Yes & Yes \\
    Year FE    & Yes & Yes & Yes & Yes \\
    \bottomrule
    \end{tabular}}
    \par\smallskip\noindent\footnotesize\textit{Notes:} Robust SE, clustered by country. $^{***}\,p<0.01$; $^{**}\,p<0.05$; $^{*}\,p<0.10$. Reference categories: consumption/expenditure; per-capita scale. WID excluded from all columns (no welfare/scale metadata). Col.~(1) is the baseline full-sample estimate; col.~(2) additionally excludes SWIID as a robustness check; cols.~(3)--(4) re-estimate the baseline separately on pre-2000 and post-2000 observations to assess period heterogeneity. All coefficients significant at 1\%. Income premia increase substantially post-2000 across all income concepts.
\end{table}

\clearpage
\begin{table}[p]
    \centering
    \small
    \caption{Practical Correction Factors: Adjusting Gini Estimates Across Welfare Concepts\label{tab:correction}}
    \adjustbox{max size={\textwidth}{0.82\textheight}}{%
    \begin{tabular}{lrrrr}
    \toprule
    & \multicolumn{2}{c}{Consumption $\to$ Income (pp)} & \multicolumn{2}{c}{Net $\to$ Gross income (pp)$^{\dagger}$} \\
    \cmidrule(lr){2-3}\cmidrule(lr){4-5}
     & Mean & Median & Mean & Median \\
    \midrule
    \multicolumn{5}{l}{\textit{Panel A: By World Bank Region}} \\
    East Asia \& Pacific       & $+$4.6 (0.4)  & $+$4.3  & $+$3.1 (0.3) & $+$2.7 \\
    Europe \& Central Asia     & $+$2.7 (0.2)  & $+$2.2  & $+$3.0 (0.1) & $+$3.2 \\
    Latin America \& Caribbean & $+$8.7 (0.6)  & $+$7.9  & $+$1.0 (0.1) & $+$1.0 \\
    Middle East \& N.\ Africa  & $+$3.4 (0.3)  & $+$3.1  & $+$3.6 (0.2) & $+$4.2 \\
    North America              & $+$10.2 (0.4) & $+$11.3 & $+$3.7 (0.1) & $+$3.8 \\
    South Asia                 & $+$7.9 (0.7)  & $+$7.0  & $+$0.3 (0.2) & $+$0.4 \\
    Sub-Saharan Africa         & $+$7.3 (0.7)  & $+$6.3  & $+$2.7 (0.3) & $+$2.9$^{*}$ \\
    \textbf{Global}            & $\mathbf{+4.7}$ (0.2) & $\mathbf{+4.3}$ & $\mathbf{+2.7}$ (0.1) & $\mathbf{+2.8}$ \\
    \midrule
    \multicolumn{5}{l}{\textit{Panel B: By World Bank Income Group}} \\
    High income                & $+$3.0 (0.2) & $+$2.4 & $+$3.2 (0.1) & $+$3.4 \\
    Upper middle income        & $+$5.3 (0.3) & $+$4.9 & $+$1.0 (0.1) & $+$1.2 \\
    Lower middle income        & $+$6.5 (0.4) & $+$5.8 & $+$0.6 (0.2) & $+$0.4 \\
    Low income                 & $+$7.7 (1.3) & $+$6.8 & \multicolumn{2}{c}{n/a} \\
    \midrule
    \multicolumn{5}{l}{\textit{Panel C: From regression (country and year fixed effects)}} \\
    Net disposable income vs.\ consumption  & $+$3.7 (0.5) & --- & \multicolumn{2}{c}{---} \\
    Gross income vs.\ consumption           & $+$5.7 (0.6) & --- & $+$2.0 & --- \\
    Mixed income vs.\ consumption           & $+$3.6 (0.6) & --- & \multicolumn{2}{c}{---} \\
    OECD-modified AE vs.\ per capita (scale) & \multicolumn{2}{c}{$-$1.2 (0.2)} & \multicolumn{2}{c}{$-$1.2 (0.2)} \\
    Household (unequalised) vs.\ per capita (scale) & \multicolumn{2}{c}{$+$2.9 (0.4)} & \multicolumn{2}{c}{$+$2.9 (0.4)} \\
    \bottomrule
    \end{tabular}}
    \par\smallskip\noindent\footnotesize\textit{Notes:} Consumption$\to$Income: matched country-year pairs with both welfare concepts available (N=1,124). Net$\to$Gross$^{\dagger}$: within-database matched country-year pairs reporting both a net and a gross income Gini (SWIID source series and UNU-WIDER; N=2,686). ``Gross'' denotes income before direct taxes and social contributions but inclusive of transfers; ``net'' denotes post-tax disposable income. Gross income is conceptually distinct from pre-tax, pre-transfer market income: these corrections capture the effect of direct taxation only, not the full market-to-disposable redistribution effect shown in Figure~\ref{fig:redistribution}. Standard errors of the means in parentheses. Panel C (regression) gross--net difference: $5.736 - 3.698 = 2.04$ pp (12 databases). Panel~C standard errors (in parentheses) are clustered by country, taken from Table~\ref{tab:metareg} column~1; the gross--net difference combines two coefficients and is reported without a separate standard error. $^{*}$Sub-Saharan Africa: N=14; interpret with caution. Low income: no matched pairs available. All values in Gini percentage points. Positive = target concept exceeds base concept.
\end{table}

\clearpage
\newpage
\section*{List of Figures}

\end{document}


\begin{center}
{\Large\textbf{Online Supplement}}\\[4pt]
{\large \textit{A World of Ginis}}\\[4pt]
Lidia Ceriani and Paolo Verme
\end{center}

\section{Database vintages and access details}\label{s:vintages}

Table~\ref{tab:s1} documents the exact vintage of each of the thirteen source
databases used to assemble the unified dataset. Because all of these databases
are revised over time (see Section~4.4 of the main text), replication requires
the specific vintages listed here.

\begin{table}[h!]
\centering
\small
\caption{Source files and vintages\label{tab:s1}}
\adjustbox{max width=\textwidth}{%
\begin{tabular}{lllp{4.6cm}}
\toprule
Database & Source file(s) & Vintage / version & Access \\
\midrule
UNU-WIDER WIID & \texttt{WIID\_08APR2025.dta} & WIID release of 8 April 2025 & \url{https://www.wider.unu.edu/database/wiid} \\
SWIID (source data) & \texttt{swiid\_source.csv} & SWIID source file, downloaded June 2025 & \url{https://fsolt.org/swiid/} \\
All the Ginis (ATG) & \texttt{allginis\_2019\_stata12.dta} & 2019 release & \url{https://stonecenter.gc.cuny.edu/research/all-the-ginis/} \\
World Bank PIP & \texttt{WB\_pip.csv} & Downloaded June 2025 & \url{https://pip.worldbank.org} \\
WID & \texttt{WID\_Dataset.dta}, \texttt{WID\_country\_dictionary.csv} & Downloaded July 2025 & \url{https://wid.world} \\
LIS & \texttt{LIS.dta} & Downloaded June 2025 & \url{https://www.lisdatacenter.org} \\
OECD IDD & \texttt{OECD\_gini.csv} & Downloaded May 2025 & \url{https://www.oecd.org/social/income-distribution-database.htm} \\
Eurostat & \texttt{tessi190.csv}, \texttt{tessi191.csv}, \texttt{ilc\_di12b.csv}, \texttt{ilc\_di12c.csv} & Downloaded August 2025 & \url{https://ec.europa.eu/eurostat} \\
CEPAL/ECLAC & \texttt{CEPALSTAT\_Gini.xlsx} & Downloaded June 2025 & \url{https://statistics.cepal.org} \\
SEDLAC & \texttt{SEDLAC\_Gini\_zero\_inc\_exc.csv}, \texttt{SEDLAC\_gini2\_onlyurban.csv} & Downloaded June 2025 & \url{https://www.cedlas.econo.unlp.edu.ar/wp/en/estadisticas/sedlac/} \\
IDB Soci\'ometro & \texttt{IDB\_gini.csv}, \texttt{IDB\_ginihh.csv}, \texttt{IDB\_ginihh\_phc.csv} & Downloaded June 2025 & \url{https://mydata.iadb.org} \\
ADB KIDB & \texttt{ADB\_Gini.csv} & Downloaded June 2025 & \url{https://kidb.adb.org} \\
Afristat & \texttt{Afristat.dta} & Downloaded June 2025 & \url{https://www.afristat.org} \\
\midrule
WB income groups & \texttt{CLASS\_2025\_10\_07.xlsx} & World Bank country classification, 7 Oct 2025 & \url{https://datahelpdesk.worldbank.org} \\
Country codes & \texttt{WB\_Country\_Codes.dta}, \texttt{countries-codes.csv} & --- & \\
Map boundaries & \texttt{world-administrative-boundaries} & --- & \url{https://public.opendatasoft.com} \\
\bottomrule
\end{tabular}}
\par\smallskip\noindent\footnotesize\textit{Note:} Download dates inferred from
file timestamps.
\end{table}

\section{Harmonised variable coding}\label{s:coding}

The unified dataset (\texttt{Dataset.dta}; 122,351 observations) harmonises the
following indicator variables across databases.

\paragraph{Welfare concept (\texttt{welfare}).}
1 = Income, net (post-tax disposable income);
2 = Income, net/gross (mixed or undetermined);
3 = Income, gross (pre-tax, inclusive of transfers);
4 = Consumption/expenditure;
5 = Earnings (in the SWIID, code 5 identifies the standardised \emph{market
income} series, used only for the redistribution analysis).
A more detailed variable (\texttt{welfare\_detailed}) preserves source-specific
concepts; e.g.\ Eurostat codes 600 (disposable), 603 (disposable before social
transfers, pensions included in transfers) and 604 (disposable before all
social transfers).

\paragraph{Equivalence scale (\texttt{scale}).}
1 = Per capita; 2 = Equivalised (OECD-modified or similar); 3 = Household
(no adjustment). Source-specific scales preserved in
\texttt{scale\_detailed}.

\paragraph{Coverage.}
\texttt{areacovr} = geographic coverage (1 = national/all); \texttt{popcovr} =
population coverage (1 = all population). The ``core'' sample used in the
quantitative analysis of Section~5 retains observations with national coverage
and total population (\texttt{areacovr} $\in \{1,.\}$ and \texttt{popcovr}
$\in \{1,.\}$) and drops the IDB individual-level earnings series. For the
cross-database analyses (Table~4, Table~5, Figures~8--9), each database is
further reduced to a single value per country-year by taking the \emph{median}
Gini, so that a database reporting several welfare concepts or sources for the
same country-year contributes one observation. This makes the selection
deterministic: keeping an arbitrary tied row would otherwise vary from run to
run, because about half of all database-country-year cells contain more than
one observation.

\paragraph{Country classification.}
World Bank region codes (\texttt{region\_code}) and income groups
(\texttt{income\_level}, \texttt{income\_levelname}) from the World Bank
classification of October 2025. Historical and non-classified entities
(e.g.\ Czechoslovakia, the Soviet Union, Taiwan (China), Venezuela) are
assigned manually as documented in \path{01\_Create\_MetaDataset.do}.

\section{Additional results referenced in the text}\label{s:additional}

\paragraph{Balanced-panel meta-regression (Section 5.4).} Restricting the
pre-2000 and post-2000 subsamples to the 157 countries observed in both
periods: net disposable income premium $+1.65$ (s.e.\ $0.60$) pre-2000 and
$+4.15$ (s.e.\ $0.64$) post-2000; gross income premium $+3.72$ (s.e.\ $0.60$)
and $+6.10$ (s.e.\ $0.67$).

\paragraph{Aggregate divergence trend (Section 5.1).} OLS slope $+0.033$ Gini
points per year (Newey--West s.e.\ $0.011$, 3 lags; $t \approx 2.9$; 95\% CI
$[0.010,\,0.055]$), estimated on the 1960--2023 annual series of mean
within-country-year ranges (years with at least five multi-database cells),
with each database entering at its median Gini per country-year.